\documentclass{article}

\usepackage{iclr2027_conference,times}
\newcommand{\iclrbst}{iclr2027_conference}
\iclrfinalcopy

\usepackage{amsmath}
\usepackage{amssymb}
\usepackage{graphicx}
\usepackage{placeins}
\usepackage{float}
\usepackage{booktabs}
\usepackage{tabularx}
\newcolumntype{Y}{>{\raggedright\arraybackslash}X}
\newcolumntype{P}[1]{>{\raggedright\arraybackslash}p{#1}}
\floatstyle{ruled}
\newfloat{algorithm}{tbp}{loa}
\floatname{algorithm}{Algorithm}
\usepackage{hyperref}
\hypersetup{hidelinks,
  pdftitle={PhaseMatcher: Autoregressive Phase-Set Identification with Spectral Decomposition},
  pdfauthor={Zhonglong Peng; Qiuliang Liu; Chang Chen; Geng Zhong; Qi Li; Lihong Wang; Lan Jiang; Shifeng Jin}}
\usepackage{url}

\title{PhaseMatcher: Autoregressive Phase-Set\\Identification with Spectral Decomposition}
\author{\normalfont
\begin{minipage}{\dimexpr\textwidth-2\tabcolsep\relax}
\centering
Zhonglong Peng\textsuperscript{1,2}\quad Qiuliang Liu\textsuperscript{1,2}\quad
Chang Chen\textsuperscript{1,2}\quad Geng Zhong\textsuperscript{1,2}\\[3pt]
Qi Li\textsuperscript{1,2}\quad Lihong Wang\textsuperscript{3}\quad
Lan Jiang\textsuperscript{4}\quad Shifeng Jin\textsuperscript{1,2}\\[6pt]
{\small \textsuperscript{1}Institute of Physics, Chinese Academy of Sciences\\
\textsuperscript{2}University of the Chinese Academy of Sciences\\
\textsuperscript{3}Xi'an Jiaotong University\\
\textsuperscript{4}Beijing University of Posts and Telecommunications}
\end{minipage}}

\begin{document}
\maketitle
\fancyhead[L]{Preprint}

\begin{abstract}
Recovering complete phase sets from powder X-ray diffraction (PXRD) is challenging when weak-phase peaks overlap stronger signals. A natural strategy is to identify phases iteratively, removing the contribution of each identified phase from the observed pattern before predicting the next. However, even after a phase is correctly identified, misestimating its contribution can distort the residual and cause subsequent errors. We introduce \textbf{PhaseMatcher}, an autoregressive framework for complete phase-set identification with physics-guided spectral decomposition. After each phase prediction, PhaseMatcher re-estimates the contributions of all selected phases and the residual from the original observation and all selected reference patterns, accounting for physically plausible variation between reference patterns and the corresponding phase contributions in the observation. The resulting residual guides subsequent phase identification, while a separate stopping module determines when the phase set is complete. On synthetic mixtures and controlled mixtures constructed from measured single-phase patterns, PhaseMatcher improves complete-set identification over the evaluated baselines. On PhaseMix-135K, it also estimates contributions and residuals more accurately than scalar subtraction.
\end{abstract}

\section{Introduction}
\label{sec:introduction}

Powder X-ray diffraction (PXRD) provides a fingerprint of crystalline materials. It helps identify minerals, track phase evolution in batteries, characterize catalysts, and determine whether synthesis produced the target material. For a single-phase sample, identification resembles object or speaker recognition: material identity is inferred from a characteristic signal. Multiphase samples present an additional challenge. Their reflections overlap in one pattern, strong phases can obscure weak minor phases, and the number of contributing phases is unknown. Periodic Labs' recent deployment of AI for XRD analysis highlights the importance of reliable phase identification in evaluating synthesis outcomes and guiding autonomous experimentation~\citep{szymanski2023autonomous,periodiclabs2026building}.

Recent methods address this problem through sequential identification, source decomposition, or refinement-based search. XRD-AutoAnalyzer identifies phases through probabilistic branching and profile subtraction~\citep{szymanski2021probabilistic}. XDecomposer predicts an unordered set of component patterns~\citep{gao2026xdecomposer}, while PhaseDifformer recursively decomposes mixtures with optional reference guidance~\citep{ito2026phasedifformer}. DARA and RADAR-PD evaluate phase combinations through physical refinement~\citep{fei2026dara,yadav2026radarpd}. These approaches motivate a complementary direction: learning phase-identity sequences together with a spectral explanation that is jointly estimated as the selected set grows.

The key difficulty is that correct phase identification does not guarantee accurate contribution estimation. Peak shifts, broadening, and relative-intensity changes can make an observed contribution differ from its library reference. In sequential subtraction, overestimating a contribution can remove evidence of unidentified phases, while underestimating it leaves already explained intensity in the residual. When the model selects another overlapping phase, its reference provides additional information for jointly estimating the selected contributions. This creates a feedback loop: selected phase identities provide reference patterns for explaining the original observation, and the resulting residual informs the next identity prediction.

We introduce \textbf{PhaseMatcher}, an autoregressive framework that couples phase-identity prediction with reference-conditioned spectral decomposition. It predicts each phase from the current residual and selected-phase history, then re-estimates all selected contributions using the original observation and their reference patterns. Physics-guided feature adaptation accounts for zero shift, strain, and broadening; nonnegative allocation conserves observed intensity across the selected phases and the residual. The updated residual guides the next phase prediction. A separate STOP head uses the original observation and phase history to decide when the set is complete, allowing background and noise to remain.

On PhaseMix-135K and controlled mixtures constructed from measured RRUFF patterns, PhaseMatcher improves complete-set identification over the evaluated baselines and produces more accurate residuals with slower error growth when the correct phases are supplied.

Our contributions are threefold:
\begin{itemize}
    \item \textbf{Joint phase identification and spectral decomposition.} PhaseMatcher couples history-conditioned phase prediction with reference-conditioned spectral decomposition and a separate STOP decision for phase sets of unknown size.
    \item \textbf{Contribution re-estimation and residual updating.} PhaseMatcher re-estimates the contributions of all selected phases and the residual from the original observation and all selected references as the selected set expands.
    \item \textbf{Identification and reconstruction evaluation.} On PhaseMix-135K and controlled mixtures constructed from measured RRUFF single-phase patterns, PhaseMatcher improves complete-set identification over the evaluated baselines. On PhaseMix-135K, automatic and correct-phase experiments show lower contribution and residual errors than Scalar, while the physics-guidance ablation shows lower contribution and residual MAE before the phase set is complete.
\end{itemize}

\section{Related Work}
\label{sec:related-work}

\subsection{Library search and residual matching}

\citet{oviedo2019fast} classify crystallographic dimensionality and space group from XRD patterns using physics-informed data augmentation. XQueryer learns single-structure retrieval over 100,315 candidates~\citep{cao2025xqueryer}. XRD-AutoAnalyzer handles mixtures through probabilistic branching, reference fitting and subtraction, and residual-based termination~\citep{szymanski2021probabilistic}. PhaseMatcher builds on residual matching by jointly estimating selected contributions from the original observation and their references.

\subsection{Source decomposition and identity assignment}

XDecomposer predicts an unordered set of phase contributions and activity scores without a known phase count~\citep{gao2026xdecomposer}. PhaseDifformer recursively splits a mixture into a component and a phase-reduced mixture, optionally using reference retrieval~\citep{ito2026phasedifformer}. In speech separation, VoiceFilter uses a reference utterance to condition the mask for a target speaker~\citep{wang2019voicefilter}. PhaseMatcher conditions joint PXRD decomposition on selected library references, using each predicted identity to explain the observation and the resulting residual to guide the next prediction.

\subsection{Constraint-guided search and refinement}

DARA explores hypotheses from user-provided or element-filtered references and validates them with Rietveld refinement~\citep{fei2026dara}. RADAR-PD uses elemental information or a known major phase to restrict candidates, followed by recursive Rietveld validation~\citep{yadav2026radarpd}. These methods use physical fitting to assess candidate explanations of the observation. PhaseMatcher instead incorporates reference-based physical constraints directly into phase-contribution and residual estimation, while scoring the full phase library without external candidate filtering.

\section{PhaseMatcher}
\label{sec:method}

\begin{figure}[!htbp]
\centering
\includegraphics[width=\linewidth]{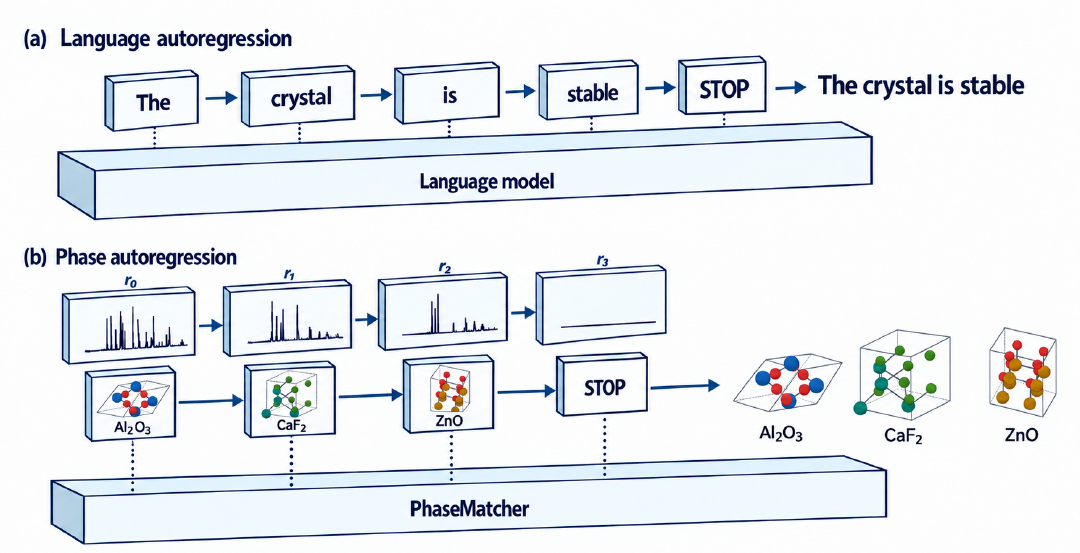}
\caption{\textbf{Autoregressive construction of a phase set.} (a) A language model predicts one token at a time until STOP. (b) PhaseMatcher predicts one phase at a time until STOP, updating the residual after each phase prediction.}
\label{fig:autoregressive-overview}
\end{figure}

PhaseMatcher constructs a phase set by alternating identity prediction and reference-conditioned contribution re-estimation (Figure~\ref{fig:autoregressive-overview}).

\subsection{Sequential phase-set identification}
\label{sec:method-problem}
\label{sec:method-decision}

Let $\mathcal L=\{0,\ldots,N-1\}$ index a library of reference patterns $\mathbf p_j\in\mathbb R_+^L$. Given an observation $\mathbf x\in\mathbb R_+^L$ normalized to unit maximum, PhaseMatcher predicts an unknown-cardinality set $Y\subseteq\mathcal L$. After $t$ selections, $H_t=(\hat y_1,\ldots,\hat y_t)$ records the ordered history and $S_t=\operatorname{set}(H_t)$ contains its distinct library IDs. Initially, $H_0=()$, $S_0=\varnothing$, and the residual query is $\mathbf r_0=\mathbf x$.

\begin{figure}[!htbp]
\centering
\includegraphics[width=\linewidth]{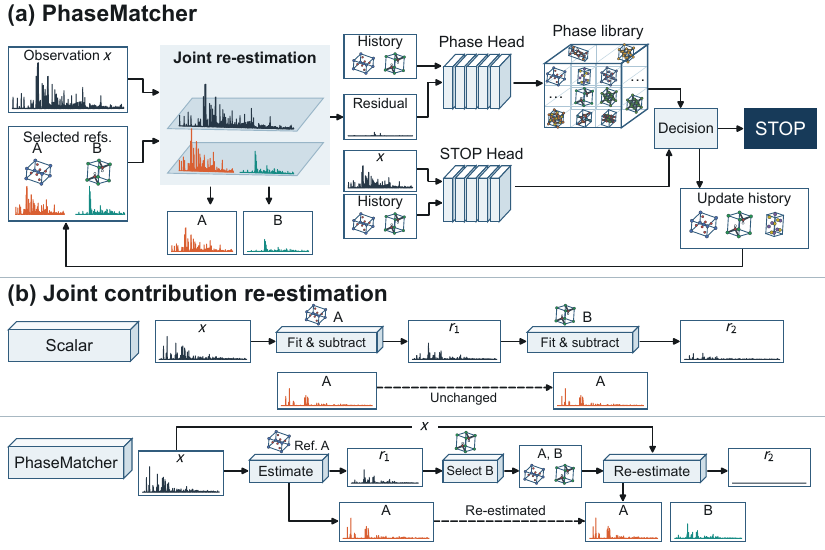}
\caption{\textbf{PhaseMatcher identifies phases while jointly re-estimating contributions.} (a) Identification and contribution re-estimation loop. (b) Sequential subtraction fixes each earlier contribution, while joint re-estimation recomputes all selected contributions.}
\label{fig:framework}
\end{figure}

The Phase Head uses $\mathbf r_t$ and $H_t$ to score the full library. The STOP Head uses the original observation $\mathbf x$ and the same history to output a STOP logit. Their logits jointly define the action distribution. This separates the spectral evidence for an additional phase from the decision that the set is complete: background and noise can remain even after every phase has been identified. Figure~\ref{fig:framework} shows the resulting loop; Appendix~\ref{app:network-config} details the shared encoder and head architectures (Figure~\ref{fig:single-step}).

A phase action appends its ID to $H_t$ and triggers residual updating. A STOP action returns $\widehat Y=S_t$; STOP has no library ID or reference pattern. For a path with $T$ phase predictions,
\begin{equation}
p(a_{0:T}\mid\mathbf x)=\prod_{t=0}^{T}p(a_t\mid\mathbf x,\mathbf r_t,H_t),
\qquad a_T=\texttt{STOP}.
\label{eq:autoregressive-path}
\end{equation}
Sequence-to-set models derive set distributions from sequence models~\citep{tang2022sequencetoset}. We use a fixed training order and evaluate the predicted identities as an unordered set.

\FloatBarrier
\subsection{Reference-conditioned contribution re-estimation}
\label{sec:method-decomposition}

\begin{figure}[!htbp]
\centering
\includegraphics[width=\linewidth]{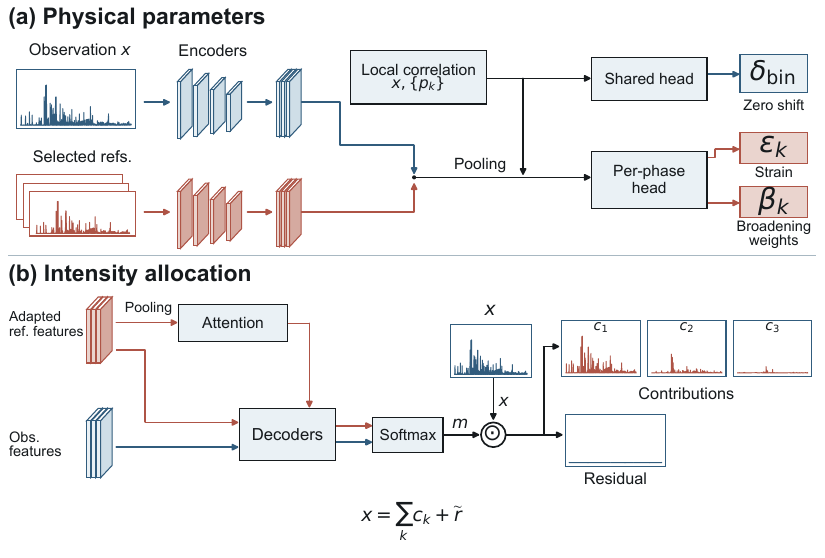}
\caption{\textbf{Reference-conditioned spectral decomposition.} (a) Physical-parameter prediction. (b) Reference-conditioned allocation of observed intensity to phase contributions and the residual.}
\label{fig:spectral-decomposition}
\end{figure}

For a nonempty history of length $t$, let $P_{S_t}=(\mathbf p_{\hat y_1},\ldots,\mathbf p_{\hat y_t})$ collect the selected references. The decomposition module computes
\begin{equation}
(\mathbf c_{1:t}^{(t)},\widetilde{\mathbf r}_t)=D_\phi(\mathbf x,P_{S_t}).
\label{eq:selected-set-decomposition}
\end{equation}
Here, $\mathbf c_k^{(t)}$ estimates the contribution of library ID $\hat y_k$ after $t$ selections. Every update starts from $\mathbf x$ and re-estimates all selected contributions $k\le t$, unlike sequential scalar subtraction (Figure~\ref{fig:framework}b). Figure~\ref{fig:spectral-decomposition} details the module.

\paragraph{Physical adaptation.}
Reference--observation variation motivates adapting reference features before intensity allocation~\citep{rietveld1969profile,cao2025simxrd}. Separate multiscale encoders process the observation and each selected reference. Local observation--reference correlations predict a shared zero shift $\delta_{\mathrm{bin}}$ in input-grid bins; correlations and pooled features predict phase-specific isotropic strain $\epsilon_k$ and broadening weights $\boldsymbol\beta_k$. At each output angle $u=2\theta$, we resample the reference features at the source coordinate
\begin{equation}
u_{\mathrm{src}}=2\arcsin\!\left[(1+\epsilon_k)\sin\!\left(\frac{u-\delta}{2}\right)\right].
\label{eq:bragg-warp}
\end{equation}
Here, $\delta=\delta_{\mathrm{bin}}\Delta u$, where $\Delta u$ is the input grid's angular spacing; all angles use radians. The broadening weights combine the resampled features with locally smoothed versions. These operations warp and smooth reference features along the angular axis; they do not modify crystal structures.

\FloatBarrier

\paragraph{Joint intensity allocation.}
Averaging the adapted reference features at the coarsest scale along the angular axis gives one token per selected phase. Self-attention exchanges information among these tokens. Contribution and residual decoders combine the resulting set context with observation features and unpooled multiscale reference features. Their outputs determine phase scores $z_{k\ell}$ and a residual score $z_{R\ell}$ at each angular bin $\ell$.

A softmax across the $t$ selected phases and the residual produces fractions
\begin{equation}
\begin{aligned}
(m_{1\ell},\ldots,m_{t\ell},m_{R\ell})&=\operatorname{softmax}(z_{1\ell},\ldots,z_{t\ell},z_{R\ell}),\\
c^{(t)}_{k\ell}&=x_\ell m_{k\ell},\qquad\widetilde r_{t\ell}=x_\ell m_{R\ell}.
\end{aligned}
\label{eq:conservative-allocation}
\end{equation}
Nonnegative fractions ensure nonnegative contributions and residuals. Their sum equals one, so allocation conserves observed intensity at every bin:
\begin{equation}
c^{(t)}_{k\ell}\geq0,\qquad 0\leq\widetilde r_{t\ell}\leq x_\ell,
\qquad x_\ell=\sum_{k=1}^{t}c^{(t)}_{k\ell}+\widetilde r_{t\ell}.
\label{eq:conservation}
\end{equation}
The next identity decision uses the max-normalized residual
\begin{equation}
\mathbf r_t=\frac{\widetilde{\mathbf r}_t}{\max(\|\widetilde{\mathbf r}_t\|_\infty,10^{-6})}.
\label{eq:decomposition-state}
\end{equation}
Appendix~\ref{app:training} specifies the network architecture and physical transformations.

\subsection{Joint training and set inference}
\label{sec:method-training}

Training proceeds through single-phase pretraining, joint phase identification and decomposition, and joint training with STOP. The first stage trains the spectrum encoder and phase-identity representations on individual reference spectra using classification cross-entropy:
\begin{equation}
\mathcal L_1=\mathcal L_{\mathrm{single}}.
\label{eq:single-objective}
\end{equation}
These parameters initialize the shared spectrum encoder and phase table for multiphase identification.

Stage two orders true phases by decreasing mixture weight; for each phase target, the history contains its preceding phases. The initial query is the observation; every later query comes from $D_\phi$ applied to that observation and the prefix references. Next-phase cross-entropy supervises histories with a remaining phase. Spectral decomposition supervision covers every nonempty prefix, including complete histories:
\begin{equation}
\mathcal L_2=\lambda_{\mathrm{phase}}\mathcal L_{\mathrm{phase}}+
\lambda_{\mathrm{decomp}}\mathcal L_{\mathrm{decomp}}.
\label{eq:phase-joint-objective}
\end{equation}
The decomposition loss constrains contribution intensity and shape, source allocation, and residual shape. Next-phase gradients also pass through the normalized residual into $D_\phi$, training the update for subsequent identification. Appendix~\ref{app:training} defines its seven weighted terms and the construction of conservative training targets.

The third stage initializes from the final second-stage checkpoint and adds the STOP Head. Incomplete histories target the next phase; complete histories target STOP. We average cross-entropy from the joint action distribution separately over phase-target and STOP-target steps, while continuing decomposition supervision:
\begin{equation}
\mathcal L_3=\lambda_{\mathrm{phase}}\mathcal L_{\mathrm{phase}}+
\lambda_{\mathrm{stop}}\mathcal L_{\mathrm{stop}}+
\lambda_{\mathrm{decomp}}\mathcal L_{\mathrm{decomp}}.
\label{eq:joint-objective}
\end{equation}
All $\lambda$ are nonnegative weights. Training samples phase candidates; inference scores the complete library. Appendix~\ref{app:training} gives initialization, optimization, and sampling settings.

\paragraph{Inference.}
Inference masks already-selected IDs and prevents STOP before the first phase. A path ends at STOP or the safety cap $K_{\max}=4$. Greedy selects the highest-scoring valid action at each step. Beam retains and expands paths by cumulative action log probability; each path independently maintains its history and recomputes its residual from the original observation. Both strategies use the same checkpoint. Appendix~\ref{app:inference-details} specifies path completion, ranking, and the masks that apply only at inference; Figure~\ref{fig:greedy-beam} illustrates the search paths.

\section{Experiments}
\label{sec:evaluation}

\subsection{Experimental setup}
\label{sec:evaluation-setup}

PhaseMix-135K contains approximately 12.14 million training and 100,000 test mixtures, with 135,258 Materials Project structure records as library candidates~\citep{jain2013materials}. RRUFF mixtures draw from a library of 740 measured single-phase references~\citep{lafuente2015rruff}, with 100,000 training and 7,400 test samples. Both benchmarks construct one- to four-phase mixtures with spectral perturbations, background, and noise. We train separately on each benchmark and test on held-out phase-level perturbations. Appendix~\ref{app:benchmarks} details construction and splits.

We compare iterative database matching, XDecomposer~\citep{gao2026xdecomposer}, PhaseDifformer~\citep{ito2026phasedifformer}, and scalar-subtraction PhaseMatcher (Scalar). The source-decomposition baselines are trained or fine-tuned on each benchmark and map generated components to identities by full-library retrieval. Scalar retains the identification architecture but fits a coefficient in $[0,1]$ for each selected reference, subtracts the scaled reference from the current residual, clips negative intensities to zero, and max-normalizes the remainder. It uses this update during both training and inference and shares the full model's single-phase initialization. Appendices~\ref{app:baselines} and~\ref{app:training} give baseline protocols and training settings.

Exact-set@1 measures complete-set identification. We also report phase recall, phase-count accuracy, and the top-ten candidate hit rate (Exact-set@10), with order-independent comparison of library IDs. All methods search the full library without being given each test mixture's phase identities or phase count. Greedy and Beam use the same checkpoint, with beam width 10. Appendix~\ref{app:evaluation-metrics} defines the metrics and candidate conventions.

\subsection{Complete phase-set identification}
\label{sec:evaluation-main}
\label{sec:evaluation-secondary}

PhaseMatcher achieves the highest complete-set accuracy among the evaluated baselines on both benchmarks (Table~\ref{tab:mild-main}). The full model improves Exact-set@1 over Scalar by 9.17/12.41 percentage points under Greedy/Beam on PhaseMix-135K and 22.62/23.20 points on RRUFF mixtures.

\begin{table}[H]
\centering\small
\caption{Complete-set identification on both benchmarks. For iterative matching, Exact-set@10 uses Beam-10; the other metrics use Greedy. A dash denotes an unreported metric.}
\label{tab:mild-main}
\label{tab:rruff-current}
\setlength{\tabcolsep}{1pt}
\begin{tabular*}{\textwidth}{@{\extracolsep{\fill}}lrrrr@{}}
\toprule
Method & Exact-set@1 (\%) & Phase recall (\%) & Count acc. (\%) & Exact-set@10 (\%) \\
\midrule
\multicolumn{5}{l}{(a) PhaseMix-135K} \\
PhaseDifformer & 4.06 & 34.90 & 44.33 & - \\
Iterative database matching & 13.85 & 49.98 & 24.39 & 15.86 \\
XDecomposer & 14.63 & 51.10 & 34.95 & 18.52 \\
\midrule
PhaseMatcher, scalar Greedy & 67.98 & 88.22 & 87.59 & - \\
PhaseMatcher, scalar Beam & 69.57 & 88.47 & 86.07 & 88.79 \\
PhaseMatcher, Greedy & 77.15 & 91.50 & 93.11 & - \\
PhaseMatcher, Beam & \textbf{81.98} & \textbf{93.59} & \textbf{94.35} & \textbf{94.73} \\
\midrule
\multicolumn{5}{l}{(b) RRUFF mixtures} \\
PhaseDifformer & 2.12 & 47.04 & 37.51 & - \\
Iterative database matching & 24.47 & 66.32 & 33.66 & 28.49 \\
XDecomposer & 15.46 & 56.52 & 35.27 & 18.11 \\
\midrule
PhaseMatcher, scalar Greedy & 56.66 & 82.97 & 65.51 & - \\
PhaseMatcher, scalar Beam & 57.39 & 83.16 & 64.77 & 78.86 \\
PhaseMatcher, Greedy & 79.28 & 92.47 & 86.35 & - \\
PhaseMatcher, Beam & \textbf{80.59} & \textbf{93.07} & \textbf{86.69} & \textbf{93.72} \\
\bottomrule
\end{tabular*}
\end{table}

From one to four phases, the Greedy advantage over Scalar grows from 1.10 to 14.54 percentage points on PhaseMix-135K and from 2.43 to 31.32 on RRUFF mixtures (Table~\ref{tab:phasemix-count}). The weakest-phase recall advantage over Scalar is largest at low intensity shares and increases with spectral overlap within the weak-phase subset (Appendix~\ref{app:weak-overlap}).

\begin{table}[H]
\centering\small
\caption{Identification by phase count on both benchmarks.}
\label{tab:phasemix-count}
\setlength{\tabcolsep}{1pt}
\begin{tabular*}{\linewidth}{@{\extracolsep{\fill}}lrrrr@{}}
\toprule
Exact-set@1 / Phase recall (\%) & 1 phase & 2 phases & 3 phases & 4 phases \\
\midrule
\multicolumn{5}{l}{(a) PhaseMix-135K} \\
PhaseDifformer & 24.18 / 72.43 & 7.77 / 47.18 & 0.28 / 31.08 & 0.01 / 22.26 \\
Iterative database matching & 54.21 / 81.22 & 26.44 / 60.43 & 8.19 / 47.92 & 1.70 / 38.49 \\
XDecomposer & 72.17 / 74.67 & 29.21 / 59.21 & 5.23 / 49.47 & 0.00 / 42.37 \\
\midrule
PhaseMatcher, scalar Greedy & 93.80 / 93.96 & 84.09 / 91.70 & 69.28 / 88.50 & 52.50 / 84.84 \\
PhaseMatcher, scalar Beam & 95.72 / 95.79 & 86.64 / 93.06 & 71.75 / 89.31 & 52.88 / 83.72 \\
PhaseMatcher, Greedy & 94.90 / 95.02 & 87.64 / 93.65 & 77.70 / 91.75 & 67.04 / 89.37 \\
PhaseMatcher, Beam & \textbf{97.43 / 97.44} & \textbf{92.23 / 95.98} & \textbf{84.67 / 94.36} & \textbf{70.98 / 90.85} \\
\midrule
\multicolumn{5}{l}{(b) RRUFF mixtures} \\
PhaseDifformer & 7.43 / 91.08 & 6.01 / 59.66 & 0.59 / 42.63 & 0.00 / 33.03 \\
Iterative database matching & 64.05 / 95.81 & 42.50 / 76.55 & 20.50 / 64.20 & 8.55 / 55.41 \\
XDecomposer & 82.03 / 84.46 & 34.32 / 66.79 & 1.26 / 53.66 & 0.03 / 46.54 \\
\midrule
PhaseMatcher, scalar Greedy & 95.95 / 97.30 & 81.55 / 90.98 & 56.22 / 82.93 & 34.73 / 75.42 \\
PhaseMatcher, scalar Beam & 97.57 / 98.65 & 84.05 / 91.96 & 55.99 / 82.73 & 35.07 / 75.20 \\
PhaseMatcher, Greedy & 98.38 / 98.92 & 92.64 / 96.52 & 81.67 / 93.27 & 66.05 / 88.22 \\
PhaseMatcher, Beam & \textbf{99.05 / 99.46} & \textbf{94.19 / 97.16} & \textbf{83.33 / 93.84} & \textbf{67.13 / 88.84} \\
\bottomrule
\end{tabular*}
\end{table}

\begin{figure}[H]
\centering
\includegraphics[width=\linewidth]{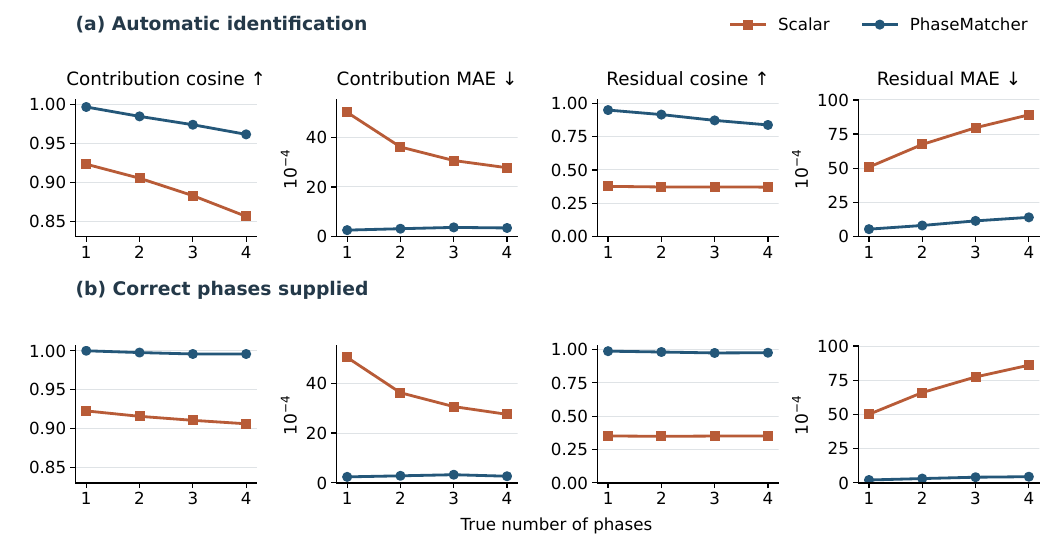}
\caption{\textbf{Contribution and residual quality by phase count.} (a) Automatic Greedy predictions at termination. (b) Both methods receive all correct phases. Residual targets subtract the correctly selected true contributions from the observation and clip at zero; both methods use the same target in (b).}
\label{fig:mechanism}
\end{figure}

\FloatBarrier
\begin{samepage}
\subsection{Contribution and residual reconstruction}
\label{sec:evaluation-decomposition}

We evaluate contribution and residual reconstruction on all 100,000 PhaseMix-135K test samples under automatic identification and with correct phases supplied. Figure~\ref{fig:mechanism} reports cosine similarity and mean absolute error (MAE) by phase count, using the observation's intensity scale (Appendix~\ref{app:automatic-residual}).
\par\end{samepage}

\paragraph{Automatic identification.}
PhaseMatcher achieves higher cosine and lower MAE for both contributions and residuals at every phase count (Figure~\ref{fig:mechanism}a). From one to four phases, terminal residual MAE rises from 0.00509 to 0.00891 for Scalar and from 0.00055 to 0.00141 for PhaseMatcher. Table~\ref{tab:decomposition-quality} compares other baselines. Scalar's decreasing absolute contribution MAE accompanies increasing relative intensity error (Appendix~\ref{app:automatic-residual}).

\paragraph{Correct phases supplied.}
With the same correct phases supplied in decreasing-weight order, PhaseMatcher achieves higher contribution and residual cosine and lower MAE than Scalar at every phase count (Figure~\ref{fig:mechanism}b). At four phases, contribution cosine is 0.9958 versus 0.9060 and residual cosine is 0.9753 versus 0.3518; joint scalar refitting remains worse on a 1,000-sample subset (Appendix~\ref{app:joint-scalar}). Across sequential updates (Figure~\ref{fig:controlled-stepwise}), residual MAE in four-phase samples rises by 2.08\% for PhaseMatcher versus 12.40\% for Scalar from the second to the third update, while cosine falls by 0.0141 versus 0.1994. Across incomplete histories, PhaseMatcher reduces residual MAE by 92.22\% relative to Scalar.

\begin{table}[!htbp]
\centering\small
\caption{Automatic contribution reconstruction on PhaseMix-135K. Beam uses the first-ranked path.}
\label{tab:decomposition-quality}
\setlength{\tabcolsep}{12pt}
\begin{tabular}{@{}lrr@{}}
\toprule
Method & Cosine similarity $\uparrow$ & MAE ($10^{-4}$) $\downarrow$ \\
\midrule
PhaseDifformer & 0.4683 & 64.6040 \\
XDecomposer & 0.6319 & 32.8429 \\
PhaseMatcher, scalar Greedy & 0.8807 & 32.4846 \\
PhaseMatcher, scalar Beam & 0.8701 & 32.5473 \\
PhaseMatcher, Greedy & 0.9735 & 3.4209 \\
PhaseMatcher, Beam & \textbf{0.9765} & \textbf{3.0280} \\
\bottomrule
\end{tabular}
\end{table}

\FloatBarrier
\begin{samepage}
\subsection{Effect of physics guidance}
\label{sec:evaluation-physical}

We compare the full model with a model retrained without physics guidance (Appendix~\ref{app:physical-guidance}). Both models receive the same correct phases sequentially and retain nonnegative, intensity-conserving allocation.
\par\end{samepage}

\begin{figure}[!htbp]
\centering
\includegraphics{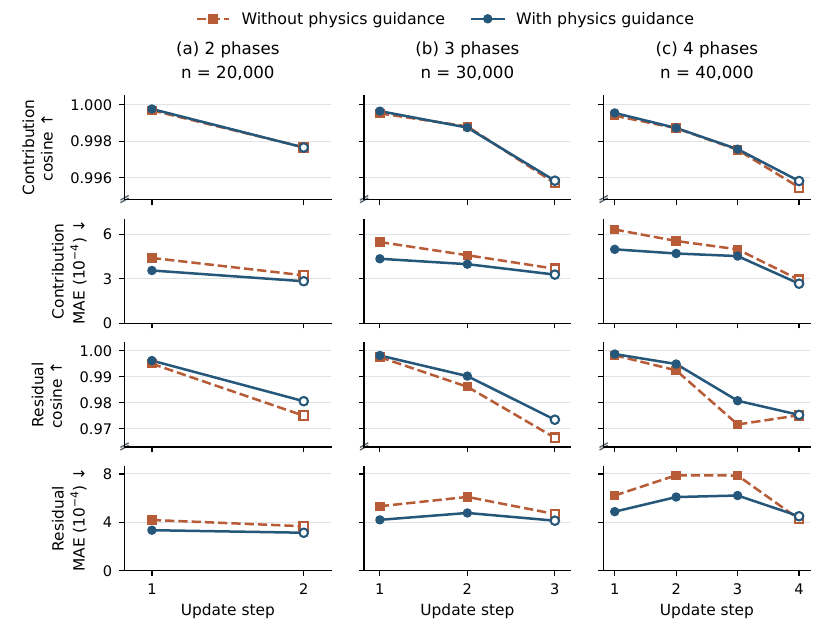}
\caption{\textbf{Physics guidance across sequential updates.} (a--c) Two-, three-, and four-phase samples with correct phases supplied. Contribution scores average the selected phases within each sample; open markers denote the complete set.}
\label{fig:physical-guidance-steps}
\end{figure}

Contribution cosine similarity differs little between models, but physics guidance lowers contribution MAE at every step (Figure~\ref{fig:physical-guidance-steps}). Before all phases are supplied, contribution and residual MAE decrease by 16.4\% and 21.6\%, respectively, improving estimates for subsequent identification.

For four-phase samples, residual MAE is 22.9\% and 21.0\% lower after the second and third updates, with higher residual cosine. After the final update, contribution MAE remains lower, but residual MAE is slightly higher and its cosine nearly unchanged. Benefits are thus clearest before completion, when residuals guide further identification. In this training comparison, Greedy and Beam also gain 0.36 and 0.62 percentage points in Exact-set@1 (Appendix~\ref{app:physical-guidance}).

\section{Conclusion}
\label{sec:conclusion}

PhaseMatcher couples sequential identification and reference-conditioned decomposition: each selected identity supplies a reference, and the updated residual guides the next prediction. It jointly re-estimates selected contributions and the residual from the original observation as the set grows. On PhaseMix-135K and RRUFF-based mixtures, the full model improves complete-set identification over Scalar. With correct phases supplied, it yields more accurate residuals with slower error growth; physics guidance improves reconstruction. Future work will study experimental transfer and reinforcement learning for phase selection and stopping.

\section*{Reproducibility Statement}

Appendix~\ref{app:benchmarks} specifies the reference libraries, mixture construction, perturbations, splits, and stored fields. Appendix~\ref{app:training} gives the architecture, losses, optimization schedules, and inference procedures; Appendix~\ref{app:baselines} documents the baseline protocols. Appendix~\ref{app:evaluation-metrics} defines the metrics and reports the complete evaluation breakdowns. We plan to release the implementation, configuration files, and data-generation and evaluation scripts after acceptance.

\FloatBarrier
\section*{AI Use Statement}

We used OpenAI tools (ChatGPT and Codex) to assist with method implementation and preprocessing code development. These tools also provided feedback on experimental design and supported result analysis and interpretation. We used generative AI to create and revise scientific illustrations, assist with literature searches, and draft, translate, and edit the manuscript. The authors reviewed all AI-assisted text, code, figures, and analyses. The authors take responsibility for the final text, code, figures, experimental results, and scientific claims.

\bibliography{reference/references}
\bibliographystyle{\iclrbst}

\appendix
\raggedbottom
\section{Datasets and data representation}
\label{app:benchmarks}

\subsection{Reference libraries and mixture splits}

PhaseMix-135K uses 135,258 structure records from a fixed Materials Project snapshot~\citep{jain2013materials}. We retain each record as a distinct candidate ID without deduplication. Multiple records can describe the same physical phase; evaluation therefore compares library IDs. We calculate Cu K$\alpha$ reflections with pymatgen~\citep{ong2013pymatgen} and render Gaussian reference profiles with full width at half maximum (FWHM) $0.10^{\circ}$. The RRUFF benchmark uses 740 measured single-phase references~\citep{lafuente2015rruff} after asymmetrically reweighted penalized least squares (arPLS) background correction~\citep{baek2015arpls}; Appendix~\ref{app:current-evaluation} gives its evaluation protocol.

Both datasets use 3,501 angular points from $10^{\circ}$ to $80^{\circ}$ at $0.02^{\circ}$ spacing. Each mixture draws $K\in\{1,2,3,4\}$ distinct library IDs and weights from Dirichlet$(1)$. The manifest orders the selected IDs by decreasing weight. Each reference ID has five independently perturbed phase profiles, termed realizations, assigned to training, validation, and test splits. Table~\ref{tab:data-splits} summarizes sample counts and the observation-realization split.

\begin{table}[H]
\centering\small
\renewcommand{\arraystretch}{1.12}
\caption{Reference libraries and mixture manifests. Realizations index observation profiles of the same library ID.}
\label{tab:data-splits}
\begin{tabularx}{\linewidth}{@{}lrrY@{}}
\toprule
Quantity & PhaseMix-135K & RRUFF mixtures & Convention \\
\midrule
Reference IDs & 135,258 & 740 & Full candidate library at inference \\
Training mixtures & 12,135,258 & 100,000 & Observation realizations 0, 1, 2 \\
Validation mixtures & 100,000 & 7,400 & Observation realization 3 \\
Test mixtures & 100,000 & 7,400 & Observation realization 4 \\
Test $K=1$ & 10,000 & 740 & One selected ID \\
Test $K=2$ & 20,000 & 1,480 & Two distinct IDs \\
Test $K=3$ & 30,000 & 2,220 & Three distinct IDs \\
Test $K=4$ & 40,000 & 2,960 & Four distinct IDs \\
\bottomrule
\end{tabularx}
\end{table}

\subsection{Phase-level and mixture-level variation}

Phase-level perturbations change peak positions, widths, and relative intensities. Mixture-level effects then apply one shared measurement state to the weighted sum. Table~\ref{tab:data-perturbations} separates these operations. The phase-level generator adds no zero shift, background, or additive noise and normalizes each output to unit maximum. For RRUFF, the same procedure adds variation to measured reference spectra; its sampled parameters describe incremental changes, not the specimen's original microstructure.

\begin{table}[H]
\centering\small
\renewcommand{\arraystretch}{1.12}
\caption{Data-construction parameters. $U[a,b]$ denotes independent uniform sampling. Peak-relative quantities use the maximum of the mixture before background and noise.}
\label{tab:data-perturbations}
\begin{tabularx}{\linewidth}{@{}P{0.15\linewidth}P{0.27\linewidth}P{0.22\linewidth}Y@{}}
\toprule
Level & Parameter & Distribution / setting & Operation \\
\midrule
Reference & Gaussian FWHM & $0.10^{\circ}$ & Clean theoretical profile rendering \\
\midrule
Phase & Wavelength in perturbation formulas & $1.5406$~\AA & Scherrer broadening and thermal attenuation \\
Phase & Crystallite size & $U[80,250]$~nm & Gaussian broadening; Scherrer width at $2\theta=45^{\circ}$ \\
Phase & Isotropic strain & $U[-0.0015,0.0015]$ & Bragg-law coordinate warp \\
Phase & Thermal root-mean-square displacement & $U[0.01,0.10]$~\AA & Angle-dependent intensity attenuation \\
Phase & Relative-intensity envelope & Six anchors from $U[-0.1,0.1]$ & Linear interpolation of multiplicative modulation \\
\midrule
Mixture & Global zero shift & $U[-0.03,0.03]^{\circ}$ & One shift for all components \\
Mixture & Extra instrumental FWHM & $U[0,0.05]^{\circ}$ & Shared Gaussian broadening \\
Mixture & Background peak ratio & $U[0,0.01]$ & Cubic Bernstein background \\
Mixture & Background shape & Four coefficients from $U[0,1]$ & Max-normalized background shape \\
Mixture & Noise standard-deviation ratio & $U[0.0008,0.0020]$ & Independent zero-mean Gaussian noise at each bin \\
\bottomrule
\end{tabularx}
\end{table}

Training uses phase realizations 0 to 2, validation uses realization 3, and testing uses realization 4. After adding measurement effects, the generator clips negative observed intensities and divides the observation by its maximum. It divides every component contribution by this same maximum, preserving their relative intensities.

Two audits quantify the PhaseMix-135K variation. Across 50,000 phase profiles from 10,000 sampled IDs, the median clean-to-perturbed cosine is 0.9148 and rises to 0.9788 after the best local shift within ten bins. Among 855,820 matched clean peaks, 97.30\% move by at most $0.1^{\circ}$. Across 10,000 realization-4 mixtures, median cosines are 0.9301 for clean versus phase-perturbed mixtures, 0.9181 for clean versus final observations, and 0.9799 for phase-perturbed mixtures versus final observations.

\subsection{Stored arrays and training-batch fields}

The pipeline stores reference banks and mixture manifests separately. A manifest specifies identities and weights; the loader generates the observed mixture and supervision tensors. Table~\ref{tab:data-files} gives the PhaseMix-135K array schema. Let $N=135{,}258$, $L=3{,}501$, $R=5$, and $M$ denote the number of mixtures in a split. Phase IDs $j\in\{0,\ldots,N-1\}$ directly index reference-bank rows, matching Section~\ref{sec:method-problem}. The array \texttt{entry.npy} maps each row to its Materials Project record ID.

\begin{table}[H]
\centering\small
\renewcommand{\arraystretch}{1.12}
\caption{PhaseMix-135K arrays on disk. Clean and perturbed banks use separate directories. A split prefix $s$ is \texttt{train}, \texttt{val}, or \texttt{test}.}
\label{tab:data-files}
\begin{tabularx}{\linewidth}{@{}P{0.31\linewidth}P{0.17\linewidth}P{0.12\linewidth}Y@{}}
\toprule
Array & Shape & Type & Meaning \\
\midrule
Clean \texttt{patterns.npy} & $N\times L$ & float32 & Candidate reference spectra \\
Perturbed \texttt{patterns.npy} & $N\times R\times L$ & float16 & Five observation profiles per ID \\
\texttt{entry.npy} & $N$ & Unicode & Record IDs in reference-row order \\
\texttt{axis\_two\_theta.npy} & $L$ & float32 & Angular grid in degrees \\
$s$\texttt{\_ids.npy} & $M\times4$ & int32 & Selected library-row IDs \\
$s$\texttt{\_weights.npy} & $M\times4$ & float32 & Mixture weights \\
$s$\texttt{\_counts.npy} & $M$ & uint8 & Number of valid components \\
\bottomrule
\end{tabularx}
\end{table}

Table~\ref{tab:data-batch} describes a joint-training batch of size $B$. For a sample with $K$ phases, valid array positions on the four-slot component axis are $0,\ldots,K-1$. Prefix index $t\in\{0,\ldots,K\}$ counts selected phases; index 0 denotes the empty history. The \texttt{targets} array stores the $K$ phase IDs at positions $0,\ldots,K-1$ and the termination marker at position $K$. It encodes \texttt{STOP} as integer $N$ solely for storage; \texttt{STOP} has no phase-library ID or reference-bank row. The loader pads unused ID and target positions with $-1$, and unused spectra and weights with zeros. Floating-point tensors use float32 and integer tensors use int64.

\begin{table}[H]
\centering\small
\renewcommand{\arraystretch}{1.12}
\caption{Joint-training batch fields. Common-scale spectra share the observation's maximum as denominator. True contributions and residuals supply supervision; the Phase Head receives a model-generated residual.}
\label{tab:data-batch}
\begin{tabularx}{\linewidth}{@{}P{0.36\linewidth}P{0.18\linewidth}Y@{}}
\toprule
Field & Batched shape & Meaning / use \\
\midrule
\texttt{mixture} & $B\times1\times L$ & Max-normalized observation \\
\texttt{phase\_ids} & $B\times4$ & True IDs in decreasing-weight order; padding $-1$ \\
\texttt{phase\_weights} & $B\times4$ & Mixture weights; padding 0 \\
\texttt{counts} & $B$ & True phase count $K$ \\
\texttt{targets} & $B\times5$ & $K$ phase IDs, a termination marker, then padding $-1$ \\
\texttt{reference\_patterns} & $B\times4\times1\times L$ & Clean references for the true history \\
\texttt{observation\_realization\_ids} & $B\times4$ & Observation realization per phase; padding $-1$ \\
\texttt{experimental\_observation\_}\newline\texttt{parameters} & $B\times8$ & Shared measurement parameters listed below \\
\texttt{mixture\_common} & $B\times1\times L$ & Same observation as \texttt{mixture} \\
\texttt{component\_contributions} & $B\times4\times1\times L$ & Weighted, perturbed phase spectra on the common scale, before prefix-specific capping \\
\texttt{residual\_patterns\_common} & $B\times5\times1\times L$ & True-prefix residual targets on the common scale \\
\texttt{residual\_patterns} & $B\times5\times1\times L$ & Independently max-normalized target copies; not the learned residual query \\
\texttt{observation\_scale} & $B\times1$ & Observation maximum before normalization \\
\bottomrule
\end{tabularx}
\end{table}

The eight measurement parameters are the global zero shift, extra instrumental FWHM, background peak ratio, noise standard-deviation ratio, and four Bernstein coefficients, in that order. Single-phase pretraining instead returns a spectrum of shape $B\times1\times L$ and an ID vector of shape $B$.

For a nonempty true history $H=(j_1,\ldots,j_t)$, let $c^*_k$ denote the stored common-scale contribution for library ID $j_k$. Here, $k\in\{1,\ldots,t\}$ indexes positions within the history. Let $s^*=\sum_{k=1}^{t}c^*_k$. At each bin, training constructs conservative targets
\begin{equation}
u^*=\min(x,s^*),\qquad
y^*_k=
\begin{cases}
(c^*_k/s^*)u^*, & s^*>0,\\
0, & s^*=0,
\end{cases}
\qquad r^*=x-u^*.
\label{eq:appendix-prefix-targets}
\end{equation}
These operations act pointwise. Capping the selected sum preserves component ratios where noise makes the nominal contribution exceed the observation. The remainder retains unselected contributions and unexplained intensity, including background and noise. The same measurement draw defines every prefix of one sample.

\FloatBarrier
\section{Model architecture and training}
\label{app:training}

\subsection{Network configuration}
\label{app:network-config}

Table~\ref{tab:network-config} specifies the spectrum encoder, prediction heads, and spectral decomposition network. The encoder and prediction heads use Transformer layers~\citep{vaswani2017attention}. For batch size $B$, spectrum length $L=3{,}501$, and $t$ selected references, decomposition receives observations of shape $B\times L$ and references of shape $B\times t\times L$; masks exclude padded references. A shared shift and phase-specific strain warp reference features; smoothing weights adapt their widths.

\begin{table}[H]
\centering\small
\renewcommand{\arraystretch}{1.12}
\caption{Network configuration. The shared spectrum encoder produces 219 tokens of width 512 from a 3,501-point spectrum. The classifier size below refers to PhaseMix-135K.}
\label{tab:network-config}
\begin{tabularx}{\linewidth}{@{}P{0.28\linewidth}Y@{}}
\toprule
Module & Configuration \\
\midrule
Spectrum encoder: convolutions & Four stride-two layers; kernels 15, 11, 7, 5; channels 256, 512, 512, 512; projection to width 512 \\
Spectrum encoder: Transformer & Learned positional embeddings; 3 layers; 8 heads; feed-forward width 2,048; dropout 0.05 \\
Single-phase classifier & Learned attention pooling; normalized feature and classifier rows; learned logit scale; $135{,}258\times512$ class matrix \\
Phase Head / STOP Head & 6 / 2 Transformer decoder layers; width 512; 8 heads; feed-forward width 2,048; dropout 0.05 \\
\midrule
Decomposition encoders & Separate observation and reference pyramids; input channels $x$ and $\log(1+9x)/\log 10$; channels 24, 48, 96, 192 \\
Local correlation & One $8\times21$ cosine-similarity table per observation--reference pair: 8 angular regions and offsets $-10$ to $10$ bins; output $B\times t\times8\times21$ \\
Physical-parameter heads & Set-shared zero shift $\delta_{\mathrm{bin}}$, bounded by 10 input-grid bins; phase-specific isotropic strain in $[-0.005,0.005]$; convex weights over identity, 3-bin, and 7-bin smoothing kernels \\
Phase-set self-attention & Angular mean pooling of the adapted coarsest reference features gives $B\times t\times192$ tokens; 2 layers, 4 heads, no slot-position embeddings; output has the same shape \\
Contribution / residual decoders & Bottleneck block and 3 linear-interpolation upsampling steps with same-scale input fusion; widths 192, 96, 48, 24; kernel-1 outputs give contribution-allocation logits $B\times t\times L$ and residual logits $B\times L$ \\
Residual convolution block & Kernels 7 and 5; group normalization; Gaussian error linear unit (GELU); identity or kernel-1 skip projection \\
Intensity allocation & Concatenate logits to $B\times(t+1)\times L$ and apply softmax over sources at each bin \\
\bottomrule
\end{tabularx}
\end{table}

Local correlation uses $\log(1+9x)$ and partitions the input grid at indices $\lfloor rL/8\rfloor$, $r=0,\ldots,8$. Integer shifts use zero padding without wraparound. Averaging the correlation tables over valid references and angular regions gives a 21-offset vector for shared-shift prediction. The shift head adds learned corrections to the scaled correlations; softmax weights over offsets give the predicted shift.

Self-attention exchanges information across the $t$ pooled phase tokens. Their masked mean supplies a 192-dimensional set representation per sample, invariant to reference order; phase outputs permute with their references. Multiplying source fractions by the observation gives $t$ nonnegative contribution spectra and one remainder, with shapes $B\times t\times L$ and $B\times L$. Their sum conserves observed intensity pointwise, as in Equation~\ref{eq:conservation}.

\begin{figure}[!htbp]
\centering
\includegraphics[width=.97\linewidth]{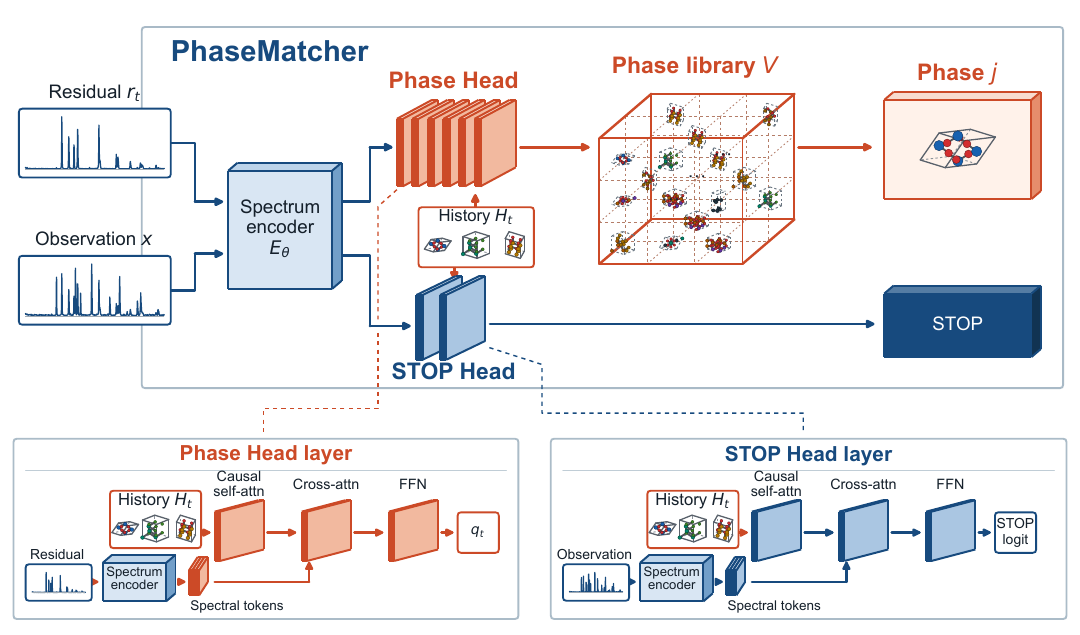}
\caption{\textbf{Dual-head network architecture.} The upper panel shows the shared encoder and each head's inputs and outputs. The lower panels expand the decoder layers into causal self-attention, cross-attention, and feed-forward blocks.}
\label{fig:single-step}
\end{figure}

\paragraph{Prediction heads.}
The two heads share the spectrum encoder, phase-identity table, and history embedding. Their history input is $[\mathrm{BOS},y_1,\ldots,y_t]$, formed from a learned beginning-of-sequence (BOS) vector and the selected phase-table rows, with learned positional embeddings. Empty history contains only BOS, which is an input token, not a phase or output action. Both decoders apply causal self-attention to this sequence. The Phase Head cross-attends to residual tokens; the STOP Head cross-attends to original-observation tokens. Each head reads its last decoder output. The Phase Head applies a learned linear projection to obtain query $h_t$; for library embedding $e_j$, its logit is
\begin{equation}
\ell_j=\frac{\langle h_t/\|h_t\|_2,e_j/\|e_j\|_2\rangle}{\tau},
\qquad \tau=0.07.
\label{eq:phase-logit}
\end{equation}
The STOP Head applies a scalar linear readout to its last decoder output. Figure~\ref{fig:single-step} shows the shared encoder and the two decoder architectures: its query $q_t$ corresponds to $h_t$ above, and $V$ denotes the phase-library representation indexed by $\mathcal L$.

\FloatBarrier

\subsection{Seven-term spectral decomposition supervision}

Both joint stages supervise every nonempty true-history prefix, including the complete set. Equation~\ref{eq:appendix-prefix-targets} defines target contributions $y_k^*$ for the $t$ selected references, their sum $u^*$, and the residual $r^*$. Let $\widehat c_k$, $\widehat u=\sum_{k=1}^{t}\widehat c_k$, and $\widehat r$ denote the predictions, with the same local index $k$. Table~\ref{tab:decomposition-loss-terms} defines the seven terms and their internal weights $\alpha_m$:
\begin{equation}
\mathcal L_{\mathrm{decomp}}=\sum_{m=1}^{7}\alpha_m\mathcal L_m.
\label{eq:decomposition-loss}
\end{equation}
The outer weight $\lambda_{\mathrm{decomp}}$ in the joint objective scales this entire sum.

\begin{table}[H]
\centering\small
\renewcommand{\arraystretch}{1.12}
\caption{Spectral decomposition loss terms and internal weights. All intensity comparisons use the common observation scale.}
\label{tab:decomposition-loss-terms}
\begin{tabularx}{\linewidth}{@{}P{0.24\linewidth}rY@{}}
\toprule
Term $\mathcal L_m$ & $\alpha_m$ & Definition \\
\midrule
Component intensity & 1 & Weighted pointwise SmoothL1 between $\widehat c_k$ and $y_k^*$; transition parameter 0.02; point weight $0.05+\sqrt{\bar y_k^*}$. Padding masks exclude absent references. \\
Component shape & 0.1 & Mean cosine distance between $\widehat c_k$ and $y_k^*$ over valid references with nonzero target norm. \\
Total selected-phase intensity & 2 & Weighted SmoothL1 between $\widehat u$ and $u^*$, with transition parameter 0.02 and point weight $0.05+\sqrt{\bar u^*}$. \\
Source allocation & 1 & Kullback--Leibler divergence from target fractions $(y_1^*,\ldots,y_t^*,r^*)/x$ to predicted fractions. Point weight $0.05+\sqrt{\bar x}+2\sqrt{\bar u^*}$; bins with $x\leq10^{-8}$ have zero weight. \\
Integrated selected-phase intensity & 0.1 & Mean $\left|\sum_\ell\widehat u_\ell-\sum_\ell u_\ell^*\right|/\max(\sum_\ell u_\ell^*,0.01)$. \\
Residual shape & 0.2 & Cosine distance between max-normalized $\widehat r$ and $r^*$. \\
Physical-deformation regularization & $10^{-4}$ & Mean over valid phases of $\tfrac12[(\delta_{\mathrm{bin}}/10)^2+(\epsilon_k/0.005)^2]+(\sum_{h=0}^{2}h\beta_{kh})^2$, with $\delta_{\mathrm{bin}}$ in input-grid bins. Kernel index $h=0,1,2$ denotes identity, 3-bin, and 7-bin smoothing, respectively. \\
\bottomrule
\end{tabularx}
\end{table}

A bar denotes max-normalization of the corresponding target spectrum. The total selected-phase intensity terms compare the summed contributions of all selected phases. Weighted pointwise terms divide the weighted error sum by the sum of point weights. Cosine terms exclude targets with norm at most $10^{-8}$ and return zero when no target qualifies. Physical-deformation regularization penalizes shift and strain away from zero and assigns a larger penalty to wider smoothing kernels. Source allocation already enforces pointwise conservation, so training adds no conservation penalty.

\subsection{Three-stage optimization}

All three stages use AdamW~\citep{loshchilov2019decoupled}; Table~\ref{tab:three-stage-training} gives the PhaseMix-135K training configuration. Stage one scores all library IDs. Each joint-training batch instead uses all positive IDs and at most 8,192 embedding-hard and random negative proposals before deduplication. Inference scores the full library.

The sampler draws 32,768 library IDs uniformly without replacement and ranks them by their maximum cosine similarity to any batch-positive ID, using detached current phase embeddings. It combines the top 4,096 with 4,096 uniformly sampled proposals and all batch-positive IDs, then removes duplicates.

With $Q$ sampled candidates, target IDs map to candidate columns $0,\ldots,Q-1$; stage three appends STOP at column $Q$. Full-library inference uses phase columns $0,\ldots,N-1$ and STOP at $N$. STOP has no phase embedding or reference spectrum.

\begin{table}[H]
\centering\small
\renewcommand{\arraystretch}{1.12}
\caption{PhaseMix-135K three-stage training. Outer loss weights are all 1. Batch-size products give GPU count, per-GPU batch size, and accumulation steps where applicable.}
\label{tab:three-stage-training}
\begin{tabularx}{\linewidth}{@{}P{0.23\linewidth}YYY@{}}
\toprule
Setting & Single-phase pretraining & Multiphase joint training & Joint training with STOP \\
\midrule
Trainable modules & Spectrum encoder and classifier & Encoder, phase table, Phase Head, decomposition & Stage-two modules and STOP Head \\
Initialization & New model & Single-phase encoder and class matrix; new Phase Head and decomposition parameters & Final stage-two weights \\
Objective & Phase cross-entropy & $\lambda_{\mathrm{phase}}\mathcal L_{\mathrm{phase}}+\lambda_{\mathrm{decomp}}\mathcal L_{\mathrm{decomp}}$ & $\lambda_{\mathrm{phase}}\mathcal L_{\mathrm{phase}}+\lambda_{\mathrm{stop}}\mathcal L_{\mathrm{stop}}+\lambda_{\mathrm{decomp}}\mathcal L_{\mathrm{decomp}}$ \\
Optimizer & AdamW & AdamW & Fresh AdamW \\
Initial learning rate & $2\times10^{-4}$ & $5\times10^{-5}$ in both parameter groups & $5\times10^{-5}$ in both parameter groups \\
Weight decay & 0.01 & $10^{-4}$ & $10^{-4}$ \\
Training duration & 500 epochs; 66,500 updates & 120,000 updates & 40,000 updates \\
Effective batch size & 1,024 ($4\times256$; no accumulation) & 126 ($7\times18$) & 224 ($7\times16\times2$ accumulation) \\
Precision & bfloat16 (BF16) mixed & BF16 mixed & BF16 mixed \\
Scheduler metric & Training loss & Validation fixed-count Exact-set@1 & Validation Greedy Exact-set@1 \\
Scheduler interval & Every epoch & Every 2,000 updates, after validation & Every 2,000 updates, after validation \\
Plateau setting & Multiply by 0.6; patience 4 & Halve after 4 non-improving validations (patience 3) & Same as stage two \\
Relative threshold & $10^{-4}$ & $10^{-4}$ & $10^{-4}$ \\
Minimum learning rate & $10^{-5}$ & $5\times10^{-6}$ & $5\times10^{-6}$ \\
Checkpoint for next stage / evaluation & Epoch 499 & Final step 120,000 & Final step 40,000 \\
\bottomrule
\end{tabularx}
\end{table}

Single-phase pretraining uses clean references and cross-entropy with label smoothing 0.05; its final learning rate is $10^{-5}$. The token encoder and class matrix initialize the shared encoder and phase table. The attention-pooling layer and logit scale are used only for pretraining.

Stage-two validation supplies each sample's true phase count $K_i$ and retrieves exactly $K_i$ phases before computing Exact-set@1. This metric monitors identification with a known count; stage-three validation uses Greedy with STOP to evaluate count prediction as well.

\subsection{RRUFF training}
\label{app:rruff-training}

PhaseMatcher initializes from RRUFF single-phase pretraining, then trains identification and decomposition jointly before adding STOP supervision. Both joint stages use the seven-term decomposition loss, AdamW with weight decay $10^{-4}$, BF16 mixed precision, and initial learning rate $5\times10^{-5}$ in both parameter groups. The learning rate halves after four non-improving validations, down to $5\times10^{-6}$. Training scores all 740 identities. The phase and STOP stages process approximately 224,000 and 160,000 training examples in total, respectively, counting repeated samples. The final STOP checkpoint supplies Greedy and Beam evaluation.

\subsection{Scalar training with subtraction-generated residuals}
\label{app:scalar-training}

Each scalar model shares its dataset's single-phase initialization and mixture splits with the corresponding decomposition model. It generates every training residual from the original observation by following the true prefix with bounded subtraction, including the complete prefix. Thus the same $[0,1]$ coefficient bound, nonnegative clipping, and query normalization apply during training and inference. Phase-stage training uses next-phase cross-entropy; STOP-stage training adds STOP cross-entropy with both outer weights equal to one. There is no decomposition network or decomposition loss.

\begin{table}[!htbp]
\centering\small
\caption{Scalar PhaseMatcher optimization. Both datasets use AdamW, weight decay $10^{-4}$, initial learning rate $5\times10^{-5}$, and BF16 mixed precision.}
\label{tab:scalar-training}
\begin{tabular*}{\linewidth}{@{\extracolsep{\fill}}lrr@{}}
\toprule
Setting & PhaseMix-135K & RRUFF mixtures \\
\midrule
Phase-stage updates / batch size & 120,000 / 128 & 1,750 / 128 \\
STOP-stage updates / batch size & 40,000 / 224 & 1,250 / 128 \\
Checkpoint for STOP / evaluation & Final phase / final STOP & Final phase / final STOP \\
\bottomrule
\end{tabular*}
\end{table}

The PhaseMix-135K scheduler uses the joint-stage settings in Table~\ref{tab:three-stage-training}. On RRUFF, the phase and STOP budgets correspond to approximately 224,000 and 160,000 sample presentations, respectively, matching those of the decomposition model.

\subsection{Joint-training and inference algorithms}
\label{app:inference-details}

Algorithm~\ref{alg:joint-training} uses true histories in decreasing-weight order, denoted $J_i=(j_{i,1},\ldots,j_{i,K_i})$ for target set $Y_i$. Next-phase cross-entropy backpropagates through the normalized residual into spectral decomposition.

Let $n(r)=r/\max(\|r\|_\infty,10^{-6})$, and let $D(x,P_H)$ return selected-phase contributions and a remainder from the original observation $x$ and all clean references $P_H$ in history $H$. Each nonempty prefix batch contributes its mean decomposition loss, weighted by its number of samples. Phase and STOP cross-entropies average their respective supervised steps separately.

\begin{algorithm}[H]
\caption{One joint-training batch (stage two or three)}
\label{alg:joint-training}
\small
\begin{tabularx}{\linewidth}{@{}rY@{}}
 & \textbf{Input:} observations $x_i$, ordered true-ID sequences $J_i$ of length $K_i$, true contributions, and stage $s\in\{2,3\}$. \\
1 & Sample one candidate set containing all batch-positive IDs and negative proposals. \\
2 & Initialize phase-loss list $A$, STOP-loss list $T$, weighted decomposition sum $d=0$, and prefix count $q=0$. \\
3 & \textbf{For} $t=0,\ldots,\max_i K_i$: \\
4 & \quad Select samples $I_t=\{i:K_i\geq t\}$; let $H_i$ contain the first $t$ IDs of $J_i$ (the empty sequence when $t=0$). \\
5 & \quad \textbf{If} $t=0$, set each residual query $r_i=x_i$. \\
6 & \quad \textbf{If} $t>0$: compute $(\widehat{\mathbf c}_i,\widehat r_i)=D(x_i,P_{H_i})$ and set $r_i=\widehat r_i$. \\
7 & \qquad Construct prefix targets using Equation~\ref{eq:appendix-prefix-targets}; compute batch loss $\ell_t$ using Table~\ref{tab:decomposition-loss-terms}. \\
8 & \qquad Accumulate $d\leftarrow d+|I_t|\ell_t$ and $q\leftarrow q+|I_t|$; \textbf{end if}. \\
9 & \quad \textbf{For each} $i\in I_t$ requiring a target ($K_i>t$ or $s=3$): \\
10 & \qquad Compute phase logits from $n(r_i)$ and $H_i$ over the sampled candidate set. \\
11 & \qquad If $s=3$, append the STOP logit from $x_i$ and $H_i$. \\
12 & \qquad If $K_i>t$, append cross-entropy for the candidate column of $j_{i,t+1}$ to $A$. \\
13 & \qquad Otherwise, append cross-entropy for STOP to $T$. \\
14 & Set $\mathcal L=\lambda_{\mathrm{phase}}\operatorname{mean}(A)+\lambda_{\mathrm{decomp}}d/q$. \\
15 & If $s=3$, add $\lambda_{\mathrm{stop}}\operatorname{mean}(T)$. \\
16 & Backpropagate $\mathcal L$ through both prediction and decomposition; update parameters after the configured gradient accumulation. \\
\end{tabularx}
\end{algorithm}

Training computes cross-entropy over the sampled phase candidates, with STOP appended in stage three. Already-selected phases remain competing classes; STOP also competes at nonterminal steps, including the empty history. These incorrect actions receive negative supervision through cross-entropy. Algorithm~\ref{alg:inference} masks repeated selections and empty-history STOP before normalizing inference probabilities.

\begin{algorithm}[H]
\caption{Greedy and Beam inference with spectral decomposition}
\label{alg:inference}
\small
\begin{tabularx}{\linewidth}{@{}rY@{}}
 & \textbf{Input:} normalized observation $x$, full reference library, maximum count $K_{\max}=4$, and width $W$ (1 for Greedy; 10 for Beam). \\
 & \textbf{State:} an ordered history $H$, its residual query $r$, and cumulative action log probability $s$. \\
1 & Initialize active paths $A=\{((),x,0)\}$ and completed paths $F=\varnothing$. \\
2 & \textbf{While} $A$ is nonempty: \\
3 & \quad Initialize next active candidates $C=\varnothing$. \\
4 & \quad \textbf{For each} $(H,r,s)\in A$: \\
5 & \qquad Compute phase logits from $r,H$ and the STOP logit from $x,H$. \\
6 & \qquad Mask IDs already in $H$; mask STOP if $|H|=0$; mask all phase actions if $|H|=K_{\max}$. \\
7 & \qquad Apply log-softmax to the remaining valid actions. \\
8 & \qquad \textbf{For each} of this path's top $W$ valid actions $a$ with log probability $\ell_a$: \\
9 & \qquad\quad If $a=\texttt{STOP}$, append $(H,r,s+\ell_a)$ to $F$. \\
10 & \qquad\quad Otherwise, append $a$ to $H$ to form $H'$; obtain remainder $\widehat r'$ from $D(x,P_{H'})$; append $(H',n(\widehat r'),s+\ell_a)$ to $C$. \\
11 & \quad Set $A$ to the top $W$ paths in $C$ and retain the top $W$ paths in $F$, each ranked separately by cumulative score. \\
12 & Return $F$ in decreasing score order. \\
\end{tabularx}
\end{algorithm}

Beam ranks paths by cumulative log probability without length normalization or unordered-set deduplication. Thus, width 10 retains up to ten completed paths, which can represent fewer than ten distinct phase sets. If Greedy's best phase and STOP logits tie, it selects the phase.

\begin{figure}[!htbp]
\centering
\includegraphics[width=.97\linewidth]{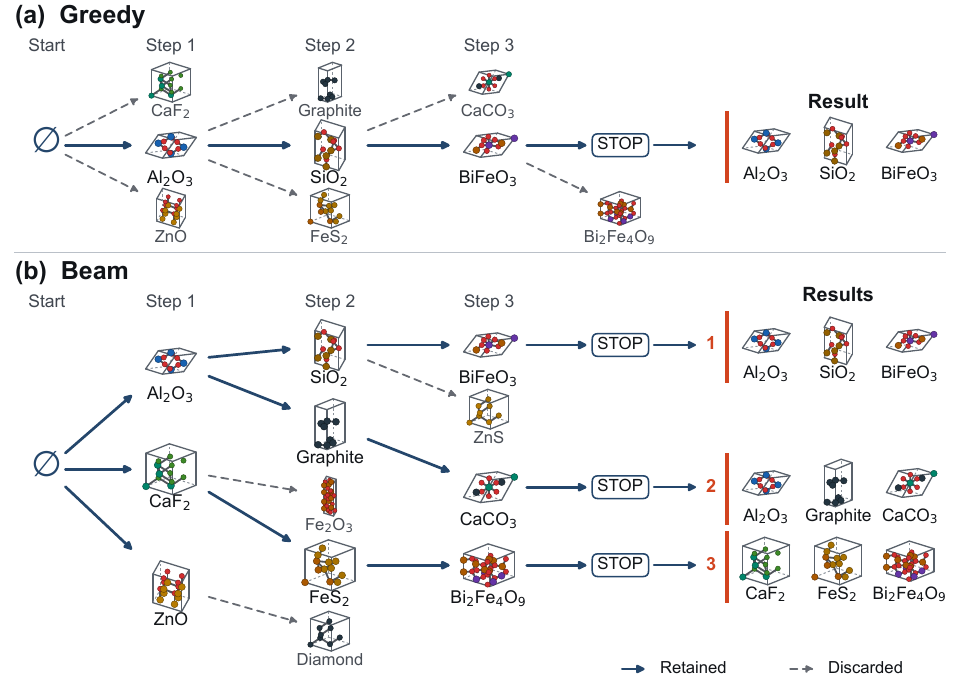}
\caption{(a) Greedy follows one path. (b) Beam expands multiple paths. Solid and dashed arrows denote retained and discarded branches; this illustration uses width three.}
\label{fig:greedy-beam}
\end{figure}

Figure~\ref{fig:greedy-beam} illustrates the search paths for Greedy and Beam.

\FloatBarrier
\section{Baseline implementations and evaluation settings}
\label{app:baselines}

\subsection{Iterative database matching}
The baseline ranks the complete reference library against the current residual by cosine similarity and selects the highest-scoring unused ID. The library contains 135,258 references for PhaseMix-135K and 740 for RRUFF. It estimates the selected reference's intensity scale by bounded nonnegative least squares (NNLS). This fit preserves peak positions, widths, and relative intensities while minimizing the pointwise squared error between the residual and the scaled reference.

\begin{samepage}
For the current residual $\mathbf r$ and nonzero reference $\mathbf p$, the coefficient has the closed-form solution
\begin{equation}
\begin{aligned}
\alpha
&=\underset{0\leq a\leq1}{\operatorname{arg\,min}}\,
\sum_{\ell}(r_\ell-a p_\ell)^2\\
&=\operatorname{clip}\left(
\frac{\langle\mathbf{r},\mathbf{p}\rangle}
{\langle\mathbf{p},\mathbf{p}\rangle},0,1\right).
\end{aligned}
\end{equation}
\end{samepage}
The upper bound of 1 is an additional constraint of this baseline. The coefficient scales spectral intensity and does not directly represent a mass or volume fraction. The baseline subtracts $\alpha\mathbf p$, clips negative intensity, and repeats for at most four phases without residual renormalization. For PhaseMix-135K, validation realization 3 selects a residual-maximum threshold of 0.22 and a next-phase cosine threshold of 0.625. For RRUFF, all 7,400 validation mixtures select thresholds of 0.12 and 0.55, respectively. Both searches use a residual-threshold grid from 0 to 0.30 in steps of 0.01 and a cosine-threshold grid from 0 to 0.90 in steps of 0.025. Validation exact-set accuracy selects the pair; count accuracy, phase recall, and smaller thresholds break ties. Each realization-4 test uses its frozen validation thresholds.

Greedy supplies this baseline's Exact-set@1, phase recall, and count accuracy in Table~\ref{tab:mild-main}; Beam-10 supplies Exact-set@10. Each Beam path maintains its own residual and expands up to ten unused references by full-library cosine similarity, with the same bounded subtraction and validation thresholds. The search always selects a first phase; later expansions require the cosine threshold. Paths finish at the residual threshold, when no candidate meets the cosine threshold, or after four phases. At each depth, the search retains up to ten unfinished paths by cumulative cosine. A separate pool ranks completed paths by mean per-step cosine, retains the highest score for each unordered set, and returns up to ten unique sets. Lexicographically smaller ID tuples break ties. Any returned set matching the target counts as an Exact-set@10 hit.

\subsection{Scalar PhaseMatcher}
This comparator keeps the PhaseMatcher backbone and replaces spectral decomposition with bounded scalar subtraction. It fits the same bounded coefficient as iterative database matching and computes $\mathbf{u}_{t+1}=\max(\mathbf{r}_t-\alpha_t\mathbf{p}_{j_t},0)$, then $\mathbf{r}_{t+1}=\mathbf{u}_{t+1}/\max(\|\mathbf{u}_{t+1}\|_\infty,10^{-6})$. A reference with squared norm at most $10^{-12}$ receives coefficient zero. Each coefficient scales one reference without changing its peak positions or widths. Both PhaseMix-135K and RRUFF models construct training queries with this same transition, applied to true histories ordered by mixture weight. The final phase-stage checkpoint initializes STOP training, and the final STOP checkpoint supplies both Greedy and Beam evaluation. Appendix~\ref{app:scalar-training} lists the training budgets.

\subsection{XDecomposer}
We use the official four-slot architecture with its reconstruction, activity, and mixture-consistency losses~\citep{gao2026xdecomposer}. A matched masked autoencoder pretrains on PhaseMix-135K training realizations before four-slot decomposition training. For set identification, validation Exact-set@1 selects the retrieval rule and activity threshold separately for each dataset. Both datasets select pure cosine retrieval, with thresholds 0.870 for PhaseMix-135K and 0.868 for RRUFF. A slot is active when its probability strictly exceeds the selected threshold. We retrieve the 20 highest-cosine references for each active source from the complete library and merge duplicate Top-1 identities for Exact-set@1, phase recall, and count accuracy.

For Exact-set@10, we expand active slots sequentially and sum their candidate scores. Repeated identities within a tuple collapse to one unordered set; different tuples producing the same set retain their maximum cumulative score. After each slot, only the ten highest-scoring unique sets survive. Ties use descending lexicographic order of sorted ID tuples. When no slot is active, the sole hypothesis is the empty set.

The alternative retrieval rule reranks the same 20 candidates by $(0.5s_{\mathrm{cos}}+0.5s_{\mathrm{diff}})g$, where $s_{\mathrm{diff}}$ is first-difference cosine. The gate $g$ inspects the candidate's three strongest bins: a bin is valid if it is below 10\% of the candidate maximum or has predicted-source support above 5\% of the source maximum within five bins. The gate passes when the valid-bin count is at least the smaller of two and the number of strong bins. This rule and the released activity threshold 0.8 are included in the validation comparison and fixed-configuration test controls in Appendix~\ref{app:xdecomposer-checks}.

\textit{Training and checkpoint selection.} The PhaseMix-135K masked autoencoder completes 2,000 epochs, with its checkpoint selected by validation reconstruction loss. The decomposition stage freezes this encoder and trains for 106,000 optimizer updates. RRUFF instead initializes from an encoder and decomposition network pretrained on clean Materials Project mixtures, then fine-tunes all parameters on the augmented RRUFF training set. Table~\ref{tab:xdecomposer-training} gives both decomposition-training budgets. Each uses a fresh AdamW optimizer, linear warmup followed by cosine decay, 32-bit floating-point precision (FP32), and an exponential moving average (EMA) with decay 0.999. Validation loss selects the EMA checkpoint; the selected checkpoints occur at the final update in both runs.

\begin{table}[!htbp]
\centering\small
\caption{XDecomposer decomposition training and RRUFF fine-tuning.}
\label{tab:xdecomposer-training}
\setlength{\tabcolsep}{6pt}
\begin{tabular}{@{}lrr@{}}
\toprule
Setting & PhaseMix-135K & RRUFF mixtures \\
\midrule
Optimizer updates & 106,000 & 13,380 (30 epochs) \\
Global batch size & 1,024 & 224 \\
Peak learning rate & $2\times10^{-4}$ & $5\times10^{-5}$ \\
Warmup updates & 2,120 & 429 \\
Encoder & Frozen & Fine-tuned \\
Selected update & 106,000 & 13,380 \\
\bottomrule
\end{tabular}
\end{table}

\subsubsection{Configuration selection and retrieval checks}
\label{app:xdecomposer-checks}

\paragraph{Validation-based configuration selection.}
We compare cosine retrieval and the reranking above on all 100,000 PhaseMix-135K and 7,400 RRUFF validation observations. The search covers 181 activity thresholds: $[0,1]$ in increments of 0.01, augmented by $[0.75,0.95]$ in increments of 0.002. Validation Exact-set@1 selects the retrieval rule and threshold; ties favor proximity to 0.8, then cosine retrieval. The selected configuration is fixed for all test identification results.

\begin{table}[!htbp]
\centering\small
\caption{XDecomposer configuration comparison on the validation sets (\%).}
\label{tab:xdecomposer-validation}
\setlength{\tabcolsep}{4pt}
\begin{tabular*}{\linewidth}{@{\extracolsep{\fill}}llrrrr@{}}
\toprule
Dataset & Retrieval & Threshold & Exact-set@1 & Phase recall & Count acc. \\
\midrule
PhaseMix-135K & Reranked & 0.800 & 11.015 & 46.180 & 46.392 \\
 & Cosine & 0.800 & 13.904 & 53.340 & 46.179 \\
 & Cosine & 0.850 & 14.572 & 51.850 & 38.569 \\
 & Cosine & 0.870 & 14.680 & 51.101 & 34.952 \\
 & Reranked & 0.870 & 11.652 & 44.232 & 35.092 \\
\midrule
RRUFF & Cosine & 0.800 & 14.608 & 59.303 & 44.986 \\
 & Cosine & 0.850 & 15.378 & 57.052 & 38.000 \\
 & Cosine & 0.868 & 15.608 & 56.202 & 34.784 \\
 & Reranked & 0.866 & 12.797 & 49.909 & 35.514 \\
\bottomrule
\end{tabular*}
\end{table}

\paragraph{Fixed-configuration test controls.}
Table~\ref{tab:xdecomposer-test-options} compares the validation-selected configurations with fixed threshold-0.8 controls using the same checkpoints and test observations. Validation-based selection improves test Exact-set@1 on both datasets, while count accuracy decreases.

\begin{table}[!htbp]
\centering\small
\caption{XDecomposer test configurations (\%). Selected rows supply the main identification tables.}
\label{tab:xdecomposer-test-options}
\label{tab:rruff-retrieval-options}
\setlength{\tabcolsep}{2pt}
\begin{tabular*}{\linewidth}{@{\extracolsep{\fill}}llrrrr@{}}
\toprule
Dataset & Retrieval, threshold & Exact-set@1 & Phase recall & Count acc. & Exact-set@10 \\
\midrule
PhaseMix-135K & Reranked, 0.800 & 10.87 & 46.16 & 46.22 & 14.15 \\
 & Cosine, 0.870 (selected) & 14.63 & 51.10 & 34.95 & 18.52 \\
\midrule
RRUFF & Cosine, 0.800 & 14.49 & 59.54 & 44.47 & 17.36 \\
 & Cosine, 0.868 (selected) & 15.46 & 56.52 & 35.27 & 18.11 \\
 & Reranked, 0.800 & 12.07 & 52.73 & 45.00 & 16.86 \\
\bottomrule
\end{tabular*}
\end{table}

\paragraph{Component retrieval versus complete-set recovery.}
For a component-level diagnostic, we align predicted slots to true contribution spectra using XDecomposer's permutation-invariant matching, retrieve each matched source against all 740 RRUFF references by cosine similarity, and average retrieval hits over true components within each sample. This diagnostic omits activity filtering, as does the released RRUFF component evaluator. The automatic phase recall and Exact-set@1 columns use the validation-selected threshold 0.868 and deduplicated predicted set.

\begin{table}[!htbp]
\centering\small
\caption{RRUFF component retrieval and automatic set recovery (\%).}
\label{tab:xdecomposer-component-check}
\setlength{\tabcolsep}{4pt}
\begin{tabular*}{\linewidth}{@{\extracolsep{\fill}}lrrrr@{}}
\toprule
Phase count & Paper component Top-1 & Our component Top-1 & Phase recall & Exact-set@1 \\
\midrule
2 & $65.31\pm2.62$ & 78.41 & 66.79 & 34.32 \\
3 & $53.49\pm1.48$ & 61.17 & 53.66 & 1.26 \\
4 & $41.90\pm1.74$ & 51.96 & 46.54 & 0.03 \\
\bottomrule
\end{tabular*}
\end{table}

The original study uses 662 RRUFF profiles, five identity folds, separate models for two to four phases, and fold-specific retrieval~\citep{gao2026xdecomposer}; ours uses one model for one to four phases, 740 profiles, held-out observation realizations, and full-library retrieval. Table~\ref{tab:xdecomposer-component-check} provides component-level context across these protocols. Within our evaluation, component retrieval is substantially higher than complete-set accuracy, which requires every identity to be correct.

Two additional checks separate input compatibility and slot selection. Retrieving the true perturbed contributions directly gives 94.06\% component Top-1 and 100\% Top-10 over all 7,400 mixtures. For the reranked-retrieval control, replacing threshold-0.8 slot selection with the $K$ highest-activity slots, where $K$ is the true count, raises Exact-set@1 from 12.07\% to 17.15\%. Duplicate identities and source-retrieval errors remain possible with the correct slot count.

\FloatBarrier

\subsection{PhaseDifformer}
\label{app:phasedifformer}
We evaluate the full pipeline of unguided and reference-guided diffusion predictors and a confidence estimator~\citep{ito2026phasedifformer}, using the released network architecture at commit \texttt{afaccfff188a}. We train all three modules from random initialization separately on each dataset. All PhaseDifformer results in the main text use the coefficient-target protocol below. Table~\ref{tab:phasedifformer-adaptation} specifies its benchmark adaptations; Table~\ref{tab:phasedifformer-results} compares it with a peak-height-target protocol.

We use coefficient targets to follow the released implementation's representation of component spectra through Gaussian basis coefficients~\citep{ito2026phasedifformer}. For our perturbed spectra, we fit nonnegative coefficients after peak filtering and select each dataset's Gaussian width using remaining-spectrum relative $L_2$ error on validation samples. Target fitting and recursive spectrum reconstruction share this basis.

The predictors minimize noise mean squared error over valid peaks. Confidence training freezes their validation-selected checkpoints and minimizes blended-prediction noise error plus $0.005$ times candidate binary cross-entropy (BCE), over all five retrieved candidates. Adam uses a global batch of 256, 426 warmup updates, and gradient clipping at 1.0. Predictor learning rates follow cosine decay after warmup; the confidence learning rate then remains constant. Diffusion loss on a fixed, phase-count-balanced validation subset of 512 mixtures selects each raw, non-EMA checkpoint (Table~\ref{tab:phasedifformer-training}).

\begin{table}[H]
\centering
\caption{PhaseDifformer training. Budgets apply to each dataset; the last column lists selected steps for PhaseMix-135K / RRUFF. All intervals count optimizer updates.}
\label{tab:phasedifformer-training}
\small
\setlength{\tabcolsep}{4pt}
\begin{tabular*}{\textwidth}{@{\extracolsep{\fill}}lrrrr@{}}
\hline
Module & Updates & Peak LR & Val. interval & Selected step \\
\hline
Unguided predictor & 213,000 & $1.5\times10^{-4}$ & 2,000 & 142,000 / 78,000 \\
Guided predictor & 213,000 & $1.5\times10^{-4}$ & 2,000 & 78,000 / 98,000 \\
Confidence estimator & 7,100 & $10^{-5}$ & 500 & 7,100 / 7,100 \\
\hline
\end{tabular*}
\end{table}

\textit{Coefficient targets.} Training samples mixture manifests with replacement and uses the realization split in Appendix~\ref{app:benchmarks}. Peaks of the observed mixture with normalized height at least 0.01 define the Gaussian basis locations. We fit nonnegative Gaussian coefficients to each true perturbed contribution. We also fit a nuisance component to the signed difference between the observation and the sum of true contributions. The fast iterative shrinkage-thresholding algorithm (FISTA) uses 2,000 iterations and a normalized projected-gradient tolerance of 0.002. Both observed peak-height conditions and coefficient targets use the observed dense-spectrum maximum as their scale.

Each predictor-training example randomly selects one true contribution for removal. The remaining-pattern target sums the other phase coefficients and the nuisance coefficients; single-phase examples retain the total coefficient target. Additional peak-position and relative-height perturbations have standard deviations $0.1^\circ$ and 0.1, respectively, applied after label fitting without refiltering the peak support. Confidence training masks each true library match with probability 0.25 before retrieving five candidates. Each candidate uses a remaining-pattern target that removes its matched true phase, or the randomly selected phase when there is no match; single-phase targets again retain the total. Validation applies no candidate dropout.

\begin{table}[!htbp]
\centering\small
\renewcommand{\arraystretch}{1.1}
\caption{PhaseDifformer coefficient-target protocol and benchmark adaptations.}
\label{tab:phasedifformer-adaptation}
\begin{tabularx}{\linewidth}{@{}P{0.23\linewidth}Y@{}}
\toprule
Item & Setting \\
\midrule
Observation generation & Seeds: 20260826 for PhaseMix-135K and 20260909 for RRUFF. PhaseMix-135K shares mixture identities and weights across baselines; differing observation seeds produce different noise realizations. \\
Gaussian representation & Filter-first NNLS targets; FWHM $0.125^\circ$ on PhaseMix-135K and $0.15^\circ$ on RRUFF, shared by target fitting and inference rendering. \\
Normalization & Observed dense-spectrum maximum, rather than a ground-truth coefficient scale. Remaining coefficients are rendered before forming subsequent peak-height conditions. \\
Peak augmentation & Position standard deviation $0.1^\circ$ follows the paper, rather than the released configuration's $1.0^\circ$; relative-height standard deviation is 0.1. \\
Confidence objective & Noise error plus $0.005$ candidate BCE over all five candidates; the released configuration uses zero diffusion weight and its top-candidate path. \\
Diffusion and clipping & The denoising diffusion probabilistic model (DDPM) uses all 1,000 indices for training and sampling, $0,\ldots,999$. Generated coefficients are nonnegative, without a peak-height upper bound. \\
Recursive filtering & Apply the 0.01 threshold on the original normalized observation scale before per-step normalization; the retained peak support only shrinks. \\
Stopping reference & NNLS-representable observation; final component amplitudes instead fit the actual observation. \\
Evaluation output & Full-library retrieval maps generated sources to a phase set for identification metrics. The original paper evaluates decomposition recovery. \\
\bottomrule
\end{tabularx}
\end{table}

\textit{Inference and set output.} Peak-position matching retrieves 15 candidate guides from the complete library. The highest candidate confidence blends guided and unguided noise predictions; confidence below 0.5 disables guidance. Sampling iterates indices 999 through 0, with fixed $\beta_t$ variance and no added noise at index 0. We restore the predicted remaining coefficients to the current scale, subtract them from the current coefficients, and clip negative values to zero to obtain the extracted-component coefficients. Rendering the remaining coefficients supplies the next peak-height condition.

Let $\bar x$ denote the NNLS reconstruction of the observation on its detected-peak basis. At each recursive step, $Q$ contains previously extracted spectra and the newly predicted remaining spectrum. The stopping error is
\begin{equation}
e_{\mathrm{stop}}=\frac{\|Qw^*-\bar x\|_2}{\max(\|\bar x\|_2,10^{-12})},
\qquad w^*=\underset{w\geq0}{\operatorname{arg\,min}}\|Qw-\bar x\|_2^2.
\end{equation}
Recursion stops when the extracted coefficient maximum is below 0.01 or $e_{\mathrm{stop}}<0.1$, accepting the predicted remainder as the final component. It otherwise continues, retaining at most four components. Final NNLS amplitude fitting compares the generated spectra with the actual observation and removes zero-norm outputs. Full-library cosine retrieval assigns each nonzero source its top-ranked identity; deduplication yields one predicted set. Contribution scores use these generated spectra, not their retrieved references. Evaluation uses one diffusion sample per observation without being given the mixture's phase identities or phase count; Exact-set@10 remains unevaluated.

\textit{Protocol comparison.} Table~\ref{tab:phasedifformer-results} also retains a peak-height-target protocol with FWHM $0.3^\circ$, peak-height output clipping, and sampling indices $999,\ldots,1$. The protocols additionally differ in augmentation, normalization, and recursive preprocessing. This comparison concerns complete protocols, not an isolated target-representation ablation.

\begin{table}[!htbp]
\centering\small
\caption{PhaseDifformer protocol comparison. The coefficient-target protocol supplies all main-text results.}
\label{tab:phasedifformer-results}
\setlength{\tabcolsep}{2pt}
\begin{tabular*}{\linewidth}{@{\extracolsep{\fill}}llrrr@{}}
\toprule
Dataset & Protocol & Exact-set@1 (\%) & Phase recall (\%) & Count acc. (\%) \\
\midrule
PhaseMix-135K & Peak height & 4.84 & 29.09 & 38.21 \\
 & Coefficient (main) & 4.06 & 34.90 & 44.33 \\
RRUFF mixtures & Peak height & 8.27 & 42.51 & 27.46 \\
 & Coefficient (main) & 2.12 & 47.04 & 37.51 \\
\bottomrule
\end{tabular*}
\end{table}

\FloatBarrier
\section{Supplementary evaluation}
\label{app:current-evaluation}

\subsection{Evaluation metrics}
\label{app:evaluation-metrics}

\paragraph{Phase-set identification.}
For $M$ test samples, let $Y_i$ be the nonempty target set and $\widehat Y_i^{(b)}$ the $b$th ranked predicted set. Each set contains library record IDs and ignores prediction order. Let $h_i\leq10$ be the number of returned candidates, and let $\mathbf{1}[\cdot]$ denote an indicator. We compute
\begin{equation}
\begin{aligned}
\mathrm{Exact\text{-}set@1}
&=\frac{1}{M}\sum_{i=1}^{M}\mathbf{1}[\widehat Y_i^{(1)}=Y_i],\\
\mathrm{Exact\text{-}set@10}
&=\frac{1}{M}\sum_{i=1}^{M}\mathbf{1}[\exists\,b\leq h_i:\widehat Y_i^{(b)}=Y_i],\\
\mathrm{Phase\ recall}
&=\frac{1}{M}\sum_{i=1}^{M}\frac{|\widehat Y_i^{(1)}\cap Y_i|}{|Y_i|},\\
\mathrm{Count\ accuracy}
&=\frac{1}{M}\sum_{i=1}^{M}\mathbf{1}[|\widehat Y_i^{(1)}|=|Y_i|].
\end{aligned}
\label{eq:set-metrics}
\end{equation}
An empty candidate list returns the empty set and fails Exact-set@10. All metrics report percentages and average over the relevant samples. Appendix~\ref{app:baselines} specifies candidate construction.

\paragraph{Contribution-pattern reconstruction.}
Predicted and target contributions use the observation's maximum intensity as their common normalization denominator. Individual components retain their relative amplitudes. For two length-$L$ patterns $a$ and $b$, define
\begin{equation}
S(a,b)=
\begin{cases}
\operatorname{clip}\!\left(\dfrac{\langle a,b\rangle}{\|a\|_2\|b\|_2},0,1\right),
& \|a\|_2\|b\|_2>10^{-16},\\
0,&\text{otherwise},
\end{cases}
\qquad
E(a,b)=\frac{1}{L}\sum_{\ell=1}^{L}|a_\ell-b_\ell|.
\label{eq:pattern-metrics}
\end{equation}
Here, $S$ measures spectral shape agreement and $E$ is mean absolute error (MAE). For sample $i$ with $\widehat K_i$ predicted and $K_i$ true components, we pad the shorter list with zero patterns to length $n_i=\max(\widehat K_i,K_i)$. Denote the padded predictions and targets by $a_{ij}$ and $b_{ij}$. A maximum-weight one-to-one assignment selects
\begin{equation}
\pi_i^*=\underset{\pi\in\mathfrak S_{n_i}}{\operatorname{arg\,max}}
\sum_{j=1}^{n_i}S(a_{ij},b_{i,\pi(j)}),
\label{eq:component-assignment}
\end{equation}
where $\mathfrak S_{n_i}$ contains all permutations of $n_i$ indices. The evaluator solves this assignment with \texttt{scipy.optimize.linear\_sum\_assignment} on the negative cosine matrix, independently of library IDs. Both metrics use this same assignment:
\begin{equation}
\begin{aligned}
\mathrm{Contribution\ cosine}
&=\frac{1}{M}\sum_{i=1}^{M}\frac{1}{n_i}\sum_{j=1}^{n_i}
S(a_{ij},b_{i,\pi_i^*(j)}),\\
\mathrm{Contribution\ MAE}
&=\frac{1}{M}\sum_{i=1}^{M}\frac{1}{n_i}\sum_{j=1}^{n_i}
E(a_{ij},b_{i,\pi_i^*(j)}).
\end{aligned}
\label{eq:contribution-metrics}
\end{equation}
Zero padding includes missing and extra components in the scores. Any pair containing a zero pattern has cosine zero, including a pair of two zero patterns. Each sample contributes equally after averaging its paired components.

\paragraph{Residual reconstruction.}
Residuals are compared on the common observation scale before query normalization. MAE uses $E$ from Equation~\ref{eq:pattern-metrics}. Residual cosine uses the normalized dot product clipped to $[0,1]$: it excludes cases with target norm at most $10^{-8}$ and assigns zero when only the predicted norm is at most $10^{-8}$. Figure~\ref{fig:mechanism} averages one terminal state per sample within each phase-count group. The stepwise controlled comparison treats each observation--history pair as one case and averages within each history group. Appendices~\ref{app:automatic-residual} and~\ref{app:stepwise} define the respective targets.

\subsection{Additional identification results}
\label{app:identification-breakdown}

Table~\ref{tab:phasemix-count} reports Exact-set@1 and phase recall by phase count on both benchmarks. Table~\ref{tab:rruff-scalar-beam} adds count accuracy and Exact-set@10 for Scalar with Beam-10 on RRUFF.

\begin{table}[!htbp]
\centering\small
\caption{Additional Scalar Beam-10 metrics on RRUFF mixtures.}
\label{tab:rruff-scalar-beam}
\begin{tabular*}{\linewidth}{@{\extracolsep{\fill}}lrrrr@{}}
\toprule
Phase count & 1 & 2 & 3 & 4 \\
\midrule
Count accuracy (\%) & 98.78 & 90.41 & 64.50 & 43.65 \\
Exact-set@10 (\%) & 100.00 & 94.80 & 81.04 & 63.99 \\
\bottomrule
\end{tabular*}
\end{table}

\subsection{Automatic identification and reconstruction}

We score automatically predicted contributions on all 100,000 PhaseMix-135K test observations. PhaseMatcher uses its final selected set and the first-ranked path for Beam. Scalar's contribution at each step is the actual intensity removed after clipping, restored to the original observation scale. XDecomposer uses active source slots at threshold 0.870; PhaseDifformer uses generated spectra after amplitude fitting and zero-source removal. All sources are paired with targets by Equation~\ref{eq:component-assignment}.

Targets are the perturbed, weighted contributions from data generation, before the conservative capping used in training (Equation~\ref{eq:appendix-prefix-targets}). Table~\ref{tab:decomposition-by-count} reports phase-count groups and includes XDecomposer's threshold-0.8 control.

\begin{table}[!htbp]
\centering\small
\caption{Automatic contribution reconstruction by phase count on PhaseMix-135K.}
\label{tab:decomposition-by-count}
\setlength{\tabcolsep}{2pt}
\begin{tabular*}{\linewidth}{@{\extracolsep{\fill}}lrrrr@{}}
\toprule
Cosine / MAE ($10^{-4}$) & 1 phase & 2 phases & 3 phases & 4 phases \\
\midrule
PhaseDifformer & 0.5464 / 31.47 & 0.5061 / 49.37 & 0.4548 / 65.94 & 0.4400 / 79.50 \\
XDecomposer ($\tau=0.8$) & 0.9270 / 10.66 & 0.7428 / 24.98 & 0.6239 / 32.27 & 0.5597 / 39.50 \\
XDecomposer ($\tau=0.870$) & 0.9342 / 13.74 & 0.7323 / 27.56 & 0.6062 / 33.99 & 0.5253 / 39.40 \\
PhaseMatcher, scalar Greedy & 0.9233 / 49.99 & 0.9052 / 36.06 & 0.8828 / 30.62 & 0.8561 / 27.72 \\
PhaseMatcher, scalar Beam & 0.9239 / 50.20 & 0.9050 / 36.11 & 0.8754 / 30.64 & 0.8352 / 27.78 \\
PhaseMatcher, Greedy & 0.9969 / 2.63 & 0.9848 / 3.20 & 0.9740 / 3.73 & 0.9616 / 3.50 \\
PhaseMatcher, Beam & 0.9998 / 2.41 & 0.9902 / 2.86 & 0.9792 / 3.34 & 0.9617 / 3.03 \\
\bottomrule
\end{tabular*}
\end{table}

Table~\ref{tab:phasedifformer-reconstruction} summarizes PhaseDifformer's generated-source quality on both complete test sets using the same checkpoints and outputs as its identification results.

\begin{table}[!htbp]
\centering\small
\caption{PhaseDifformer automatic contribution reconstruction.}
\label{tab:phasedifformer-reconstruction}
\setlength{\tabcolsep}{8pt}
\begin{tabular}{@{}lrrr@{}}
\toprule
Dataset & Test mixtures & Cosine similarity $\uparrow$ & MAE ($10^{-4}$) $\downarrow$ \\
\midrule
PhaseMix-135K & 100,000 & 0.4683 & 64.6040 \\
RRUFF mixtures & 7,400 & 0.4189 & 125.7769 \\
\bottomrule
\end{tabular}
\end{table}

\FloatBarrier
\paragraph{Residual targets and phase-count comparison.}
\label{app:automatic-residual}

For automatic Greedy inference, we score each observation at termination. Given predicted set $\widehat Y_i$, target set $Y_i$, and true contribution $c_{ij}^*$ for ID $j$, the residual target is
\begin{equation}
r_i^*(\widehat Y_i)=\max\!\left(x_i-\sum_{j\in\widehat Y_i\cap Y_i}c_{ij}^*,0\right).
\label{eq:automatic-residual-target}
\end{equation}
Only correctly selected IDs are removed from this target, so unselected phases remain and the two methods may have different target residuals. With all correct phases supplied in decreasing-weight order, both targets become $\max(x_i-\sum_{j\in Y_i}c_{ij}^*,0)$. Automatic contributions use the assignment in Equation~\ref{eq:component-assignment}; supplied-phase contributions are paired by identity.

Table~\ref{tab:reconstruction-four-metrics} lists the values in Figure~\ref{fig:mechanism} for all test observations, grouped by phase count.

\begin{table}[!htbp]
\centering\small
\caption{Contribution and residual reconstruction by phase count. Each setting includes 10,000, 20,000, 30,000, and 40,000 samples for one through four phases. MAE is in units of $10^{-4}$.}
\label{tab:reconstruction-four-metrics}
\setlength{\tabcolsep}{3pt}
\begin{tabular*}{\linewidth}{@{\extracolsep{\fill}}llrrrr@{}}
\toprule
Method & Phases & \multicolumn{2}{c}{Contribution} & \multicolumn{2}{c}{Residual} \\
\cmidrule(lr){3-4}\cmidrule(l){5-6}
 & & Cosine & MAE & Cosine & MAE \\
\midrule
\multicolumn{6}{l}{Automatic Greedy, at termination} \\
Scalar & 1 & 0.9233 & 49.9854 & 0.3770 & 50.8714 \\
Scalar & 2 & 0.9052 & 36.0601 & 0.3724 & 67.4130 \\
Scalar & 3 & 0.8828 & 30.6208 & 0.3719 & 79.6533 \\
Scalar & 4 & 0.8561 & 27.7196 & 0.3714 & 89.0785 \\
PhaseMatcher & 1 & 0.9969 & 2.6302 & 0.9488 & 5.4876 \\
PhaseMatcher & 2 & 0.9848 & 3.2005 & 0.9150 & 8.1836 \\
PhaseMatcher & 3 & 0.9740 & 3.7273 & 0.8715 & 11.5505 \\
PhaseMatcher & 4 & 0.9616 & 3.5010 & 0.8374 & 14.1280 \\
\midrule
\multicolumn{6}{l}{All correct phases supplied} \\
Scalar & 1 & 0.9226 & 50.4298 & 0.3514 & 50.1147 \\
Scalar & 2 & 0.9156 & 36.2338 & 0.3487 & 66.0129 \\
Scalar & 3 & 0.9105 & 30.6628 & 0.3508 & 77.4167 \\
Scalar & 4 & 0.9060 & 27.5469 & 0.3518 & 86.0234 \\
PhaseMatcher & 1 & 1.0000 & 2.4018 & 0.9870 & 2.0867 \\
PhaseMatcher & 2 & 0.9977 & 2.8260 & 0.9806 & 3.1298 \\
PhaseMatcher & 3 & 0.9958 & 3.2736 & 0.9735 & 4.1274 \\
PhaseMatcher & 4 & 0.9958 & 2.6630 & 0.9753 & 4.4941 \\
\bottomrule
\end{tabular*}
\end{table}

\paragraph{Absolute and relative contribution errors.}
Contribution MAE averages errors on the observation scale, so it depends on component amplitudes and the number of matched or padded components. To distinguish absolute error from intensity-relative error, we also compute
\begin{equation}
\mathrm{Relative\ L1}=\frac{1}{M}\sum_i
\frac{\sum_{j=1}^{n_i}\|a_{ij}-b_{i,\pi_i^*(j)}\|_1}
{\max(\sum_{j=1}^{K_i}\|c_{ij}^*\|_1,10^{-12})}.
\end{equation}
This uses the same automatic assignment as contribution cosine and MAE. Scalar's values rise from 0.3524 to 0.3665, 0.3767, and 0.3866 across one through four phases; PhaseMatcher's are 0.0197, 0.0334, 0.0450, and 0.0487. Thus, Scalar's decreasing absolute contribution MAE does not indicate improving relative reconstruction. To assess signal left unexplained at termination, we separately compare the final residual with $\max(x_i-\sum_{j\in Y_i}c_{ij}^*,0)$ for every sample. This complete-set-target MAE is 0.007615 for Scalar and 0.000481 for PhaseMatcher; it is distinct from the selected-set residual MAE plotted in Figure~\ref{fig:mechanism}a.

\begin{figure}[!htbp]
\centering
\includegraphics[width=\linewidth]{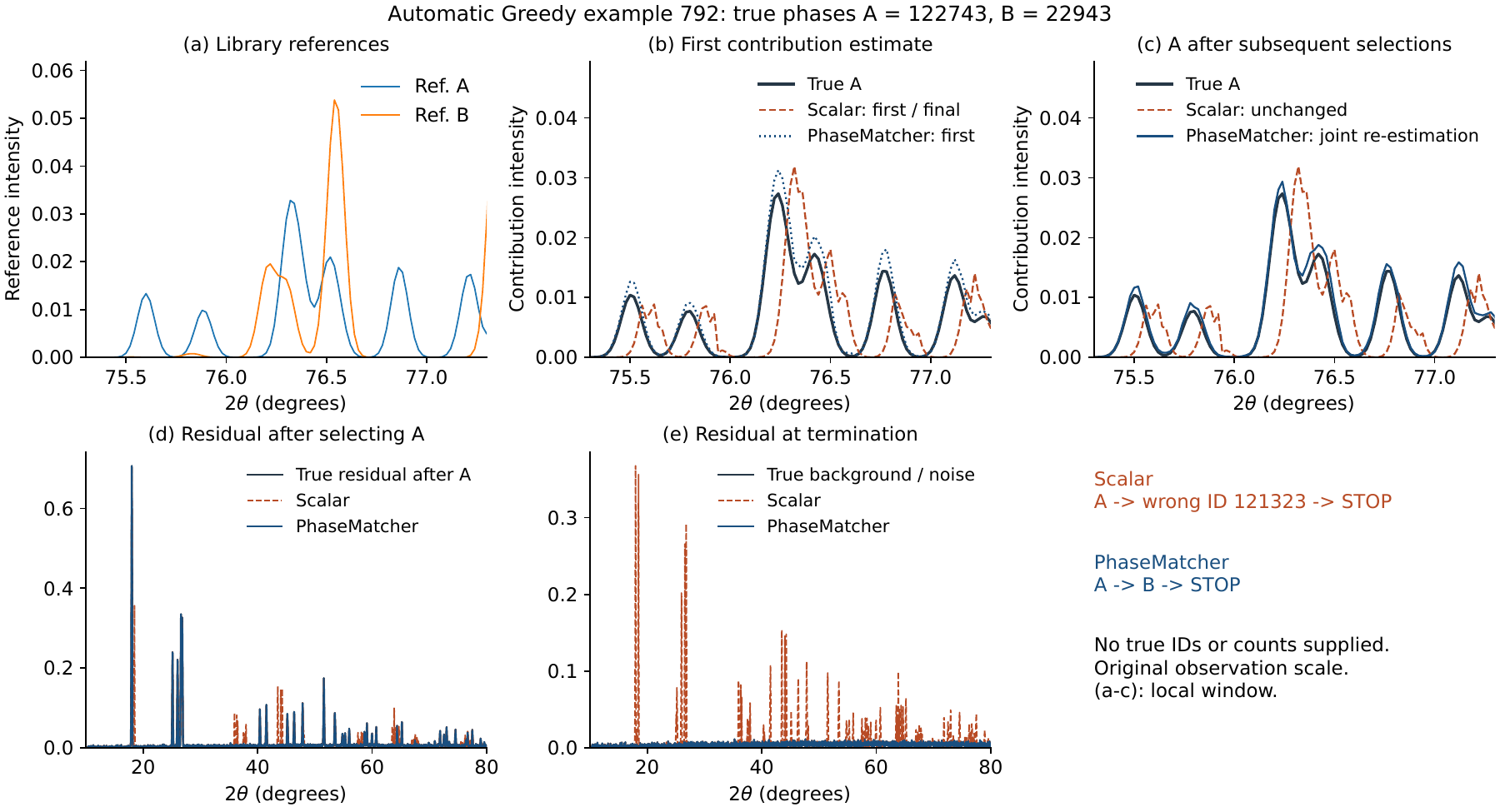}
\caption{Automatic Greedy trajectories for a selected two-phase sample. (a) Library references. (b,c) A's first and final contributions. (d) The residual after A. (e) Terminal residuals, shown against the common complete-set background/noise target. The local windows in (a--c) use unchanged reference or observation scales. Neither method receives the mixture's phase identities or phase count.}
\label{fig:automatic-revision-path}
\end{figure}

\FloatBarrier
\paragraph{An automatic trajectory with joint re-estimation.}
Figure~\ref{fig:automatic-revision-path} shows test sample 792, whose true IDs are A = 122743 and B = 22943. Both methods first select A. Scalar then selects the incorrect record 121323 and stops; PhaseMatcher selects B and stops. At $76.30^\circ$, the true A contribution is 0.017405, while Scalar removes 0.028732 and keeps this estimate fixed. After selecting B, PhaseMatcher jointly re-estimates A as 0.016523 using both references. Its full-spectrum A MAE decreases from 0.00018425 to 0.00013155, a 28.61\% reduction. The first residual supplies B's prediction.

\FloatBarrier
\subsection{Residual quality across sequential updates}
\label{app:stepwise}

We supply both methods with the correct phases in decreasing mixture-weight order for all 100,000 test observations. Evaluating every nonempty selection step gives 300,000 sample--step pairs: 200,000 before all phases are supplied and 100,000 after completion. After supplying the first $t$ phases (a prefix of length $t$), contribution targets are the actual perturbed, weighted spectra $c_k^*$, and the residual target is
\begin{equation}
r_t^*=\max\!\left(x-\sum_{k=1}^{t}c_k^*,0\right).
\label{eq:evaluation-residual}
\end{equation}
Metrics use the original observation scale before query normalization and pair contributions by identity. Scalar follows the bounded update in Appendix~\ref{app:baselines}; PhaseMatcher re-estimates all selected contributions. Complete-set residuals include background and noise.

\begin{table}[!htbp]
\centering\small
\caption{Residual reconstruction under identical true-phase histories.}
\label{tab:controlled-residual}
\setlength{\tabcolsep}{7pt}
\begin{tabular}{@{}llrr@{}}
\toprule
History & Update & Cosine similarity & MAE ($10^{-4}$) \\
\midrule
Incomplete & Scalar & 0.6634 & 65.6803 \\
Incomplete & PhaseMatcher & 0.9928 & 5.1114 \\
Complete & Scalar & 0.3508 & 75.8484 \\
Complete & PhaseMatcher & 0.9770 & 3.8705 \\
\bottomrule
\end{tabular}
\end{table}

\begin{figure}[!htbp]
\centering
\includegraphics[width=\linewidth]{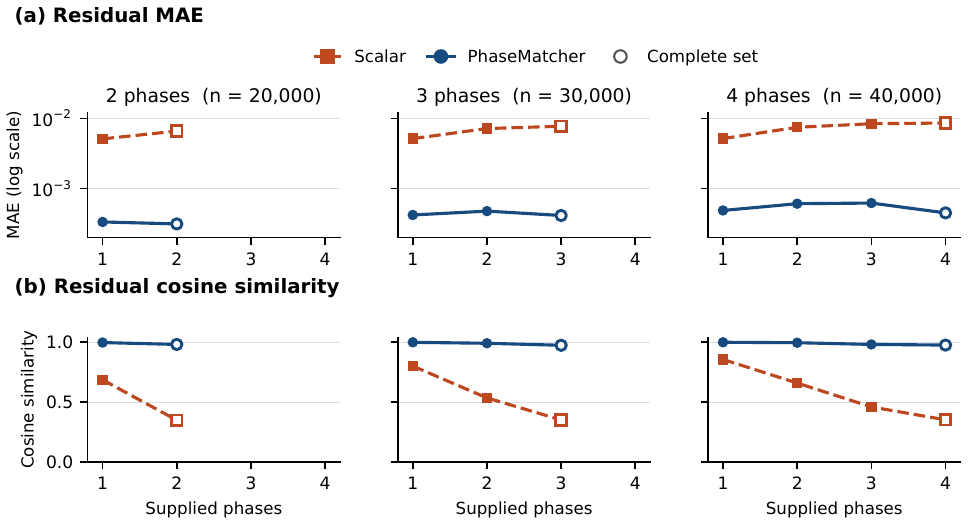}
\caption{Residual quality with the same correct phases supplied sequentially. (a) MAE on a shared logarithmic scale. (b) Cosine similarity on a shared linear scale. Each column follows a fixed group of two-, three-, or four-phase samples; points are test-set means and open markers denote the complete set.}
\label{fig:controlled-stepwise}
\end{figure}

Table~\ref{tab:stepwise-residual} and Figure~\ref{fig:controlled-stepwise} follow each phase-count group through successive updates.

Scalar's residual MAE increases at every update, while PhaseMatcher shows smaller increases at intermediate steps and lower MAE after the final update. Its residual cosine similarity also decreases less at each update. For four-phase samples, MAE grows by 45.00\% and 12.40\% across Scalar's first two transitions, compared with 24.81\% and 2.08\% for PhaseMatcher. At the final transition, Scalar's MAE rises another 2.36\%, while PhaseMatcher's falls by 27.62\%. These trends measure the quality of the residual left for subsequent identification. Individual earlier contributions need not improve monotonically: their mean error can increase at intermediate updates even though PhaseMatcher's residual remains more accurate than Scalar's.

\begin{table}[!htbp]
\centering\small
\caption{Residual reconstruction by total phase count $K$ and true-history length $t$. MAE uses the original observation scale.}
\label{tab:stepwise-residual}
\setlength{\tabcolsep}{7pt}
\begin{tabular}{@{}rrrrrrr@{}}
\toprule
$K$ & $t$ & Samples & \multicolumn{2}{c}{Scalar} & \multicolumn{2}{c}{PhaseMatcher} \\
\cmidrule(lr){4-5}\cmidrule(l){6-7}
 & & & Cosine & MAE ($10^{-4}$) & Cosine & MAE ($10^{-4}$) \\
\midrule
1 & 1 & 10,000 & 0.3514 & 50.1147 & 0.9870 & 2.0867 \\
\midrule
2 & 1 & 20,000 & 0.6855 & 51.0897 & 0.9962 & 3.3403 \\
2 & 2 & 20,000 & 0.3487 & 66.0129 & 0.9806 & 3.1298 \\
\midrule
3 & 1 & 30,000 & 0.8002 & 51.5527 & 0.9982 & 4.1955 \\
3 & 2 & 30,000 & 0.5347 & 71.7535 & 0.9902 & 4.7662 \\
3 & 3 & 30,000 & 0.3508 & 77.4167 & 0.9735 & 4.1274 \\
\midrule
4 & 1 & 40,000 & 0.8552 & 51.5652 & 0.9987 & 4.8735 \\
4 & 2 & 40,000 & 0.6586 & 74.7716 & 0.9949 & 6.0828 \\
4 & 3 & 40,000 & 0.4592 & 84.0404 & 0.9808 & 6.2091 \\
4 & 4 & 40,000 & 0.3518 & 86.0234 & 0.9753 & 4.4941 \\
\bottomrule
\end{tabular}
\end{table}

\begin{figure}[H]
\centering
\includegraphics[width=\linewidth]{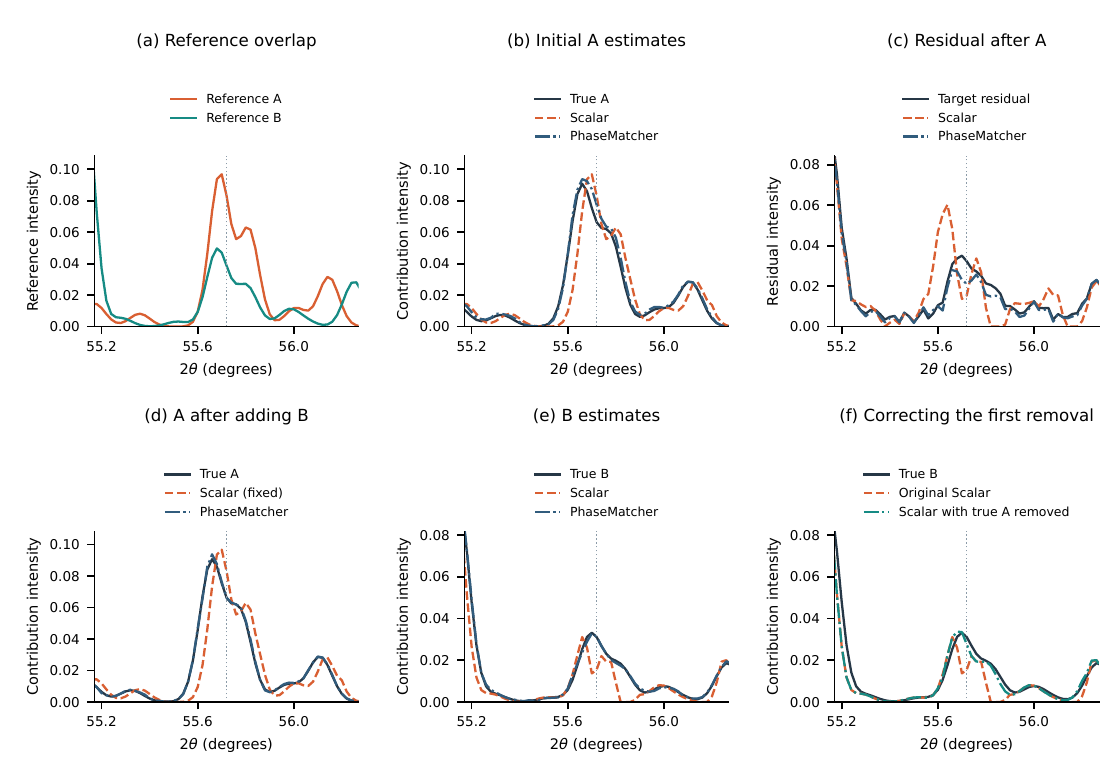}
\caption{Initial over-allocation and subsequent joint re-estimation in a selected PhaseMix sample. (a) References. (b,c) Initial A estimates and residuals. (d,e) Joint A and B estimates after adding B. (f) Scalar's B estimate when the first removal uses true A. Dotted lines mark $55.72^\circ$. References retain their library scale; contributions and residuals retain the observation scale.}
\label{fig:locked-removal-revision}
\end{figure}

\FloatBarrier
\noindent\begin{minipage}{\linewidth}
\paragraph{Joint re-estimation of an initially misallocated peak.}
Figure~\ref{fig:allocation-revision-case} illustrates a selected two-phase test sample (index 2; library IDs 44463 and 115478 for A and B). At $2\theta=26.96^\circ$, both the reference and true contribution of A are zero, but PhaseMatcher initially allocates 0.06637 intensity to A. Adding B changes this estimate to 0.00004 and yields a B contribution of 0.28088, close to its target of 0.27976. Scalar assigns zero to A at this point from the first step, but estimates B as 0.03589 after its selection. This example shows how joint re-estimation changes an initial allocation when the model adds B; Scalar already avoids that particular A error. The A contribution's full-spectrum MAE decreases from $3.0919\times10^{-4}$ to $2.3569\times10^{-4}$. These values use the original observation scale.

\begin{figure}[H]
\centering
\includegraphics[width=\linewidth]{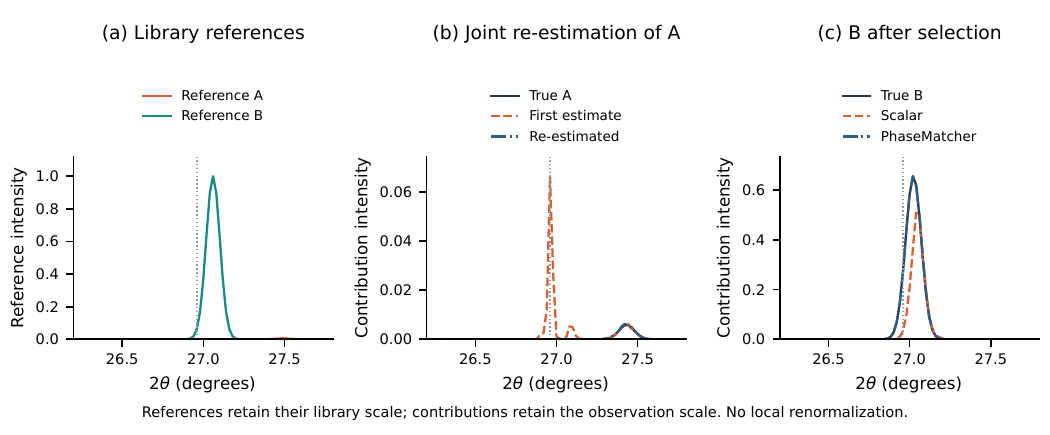}
\caption{Joint re-estimation of an initially misallocated peak in a selected PhaseMix test sample under the true history A, then A/B. (a) Reference spectra. (b) Initial and re-estimated A contributions. (c) B contributions after selection. Vertical dotted lines mark $26.96^\circ$; reference and contribution intensities retain their respective original scales.}
\label{fig:allocation-revision-case}
\end{figure}
\end{minipage}
\FloatBarrier
\clearpage
\paragraph{A fixed removal error limits the next contribution.}
Figure~\ref{fig:locked-removal-revision} shows a second selected two-phase test sample (index 60; A/B library IDs 94533/97753). Both references contain intensity near $55.72^\circ$. At this point, the true A contribution is 0.06637; Scalar and PhaseMatcher initially estimate 0.08328 and 0.07755, respectively. Both over-allocate intensity to A and reduce the signal available for B. After B joins, Scalar retains its A estimate and assigns only 0.01543 to B, whose true contribution is 0.03111. PhaseMatcher jointly re-estimates A as 0.06630 and estimates B as 0.03132. Across the full spectrum, its A MAE decreases from $5.3310\times10^{-4}$ to $4.5554\times10^{-4}$.

To isolate the effect of the first removal, we replace Scalar's removed A contribution with the true A contribution and repeat its unchanged second-step fit. The coefficient bounds, nonnegative clipping, reference B, and query normalization remain unchanged. B's estimate at $55.72^\circ$ increases to 0.02761. This intervention confirms that the initial over-removal contributes to the second-step underestimate. Both selected examples illustrate how contributions can change when the model adds another reference; Table~\ref{tab:stepwise-residual} reports residual quality over the full test set.

\FloatBarrier
\subsection{Joint scalar refitting with correct phases}
\label{app:joint-scalar}

We compare sequential scalar subtraction, joint scalar refitting, and PhaseMatcher on 1,000 PhaseMix-135K test samples, randomly selecting 250 per phase count (seed 20260916). Each method receives correct phases in decreasing-weight order, yielding 2,500 nonempty prefixes. Targets and contribution pairing follow Appendix~\ref{app:stepwise}; Table~\ref{tab:joint-scalar} averages scores over all 2,500 sample--step pairs.

Sequential subtraction uses the bounded, clipped, normalized transition in Appendix~\ref{app:baselines}. Joint scalar refitting instead returns to the original observation $x$ at every step. With selected references as columns of $P_t$, it solves
\begin{equation}
\hat w_t=\underset{0\leq w_j\leq1}{\operatorname{arg\,min}}\|P_t w-x\|_2^2,
\qquad \hat c_j=\hat w_{t,j}p_j,
\qquad \hat r_t=\max\left(x-\sum_j\hat c_j,0\right).
\label{eq:joint-scalar-fit}
\end{equation}
We use bounded-variable least squares with tolerance $10^{-10}$. All selected coefficients are refitted, but each reference retains its spectral shape.

Sequential contribution scores use the actual clipped removal. To check sensitivity to this convention, we also proportionally clip jointly fitted contributions wherever their sum exceeds the observation:
\begin{equation}
\tilde c_j(\ell)=\hat c_j(\ell)\min\left(1,\frac{x(\ell)}{\sum_q\hat c_q(\ell)}\right),
\label{eq:joint-scalar-clip}
\end{equation}
with multiplier one when the denominator is zero. This leaves the joint residual unchanged and makes $\sum_j\tilde c_j+\hat r_t=x$.

\begin{table}[!htbp]
\centering\small
\caption{Correct-phase reconstruction on the 1,000-sample subset. MAE is in units of $10^{-4}$.}
\label{tab:joint-scalar}
\setlength{\tabcolsep}{4pt}
\begin{tabular*}{\linewidth}{@{\extracolsep{\fill}}lrrrr@{}}
\toprule
& \multicolumn{2}{c}{Contribution} & \multicolumn{2}{c}{Residual} \\
\cmidrule(lr){2-3}\cmidrule(l){4-5}
Update & Cosine $\uparrow$ & MAE $\downarrow$ & Cosine $\uparrow$ & MAE $\downarrow$ \\
\midrule
Sequential scalar & 0.9202 & 40.5222 & 0.5336 & 65.0384 \\
Joint scalar & 0.8909 & 51.3037 & 0.5330 & 64.7952 \\
Joint scalar, proportional clipping & 0.9222 & 39.9904 & 0.5330 & 64.7952 \\
PhaseMatcher & \textbf{0.9983} & \textbf{3.6447} & \textbf{0.9874} & \textbf{4.2546} \\
\bottomrule
\end{tabular*}
\end{table}

Joint scalar refitting reduces residual MAE by only 0.37\% relative to sequential subtraction. Proportional clipping improves its contribution scores, but the gap to PhaseMatcher remains: PhaseMatcher has 90.89\% lower contribution MAE than the clipped joint fit and 93.43\% lower residual MAE. For four-phase samples, joint and sequential scalar residual errors remain close throughout the updates (Table~\ref{tab:joint-scalar-steps}). These results show that fitting scale coefficients alone does not recover the contribution and residual quality of reference-conditioned spectral decomposition on this subset.

\begin{table}[!htbp]
\centering\small
\caption{Residual MAE ($10^{-4}$) over the 250 four-phase samples.}
\label{tab:joint-scalar-steps}
\setlength{\tabcolsep}{8pt}
\begin{tabular}{@{}lrrrr@{}}
\toprule
Update & Step 1 & Step 2 & Step 3 & Step 4 \\
\midrule
Sequential scalar & 51.6661 & 74.1393 & 82.7750 & 84.8074 \\
Joint scalar & 51.6661 & 73.5152 & 82.2667 & 84.9592 \\
PhaseMatcher & \textbf{4.8220} & \textbf{5.8917} & \textbf{6.1350} & \textbf{4.3864} \\
\bottomrule
\end{tabular}
\end{table}
\FloatBarrier

\subsection{Physics guidance improves reconstruction}
\label{app:physical-guidance}

Both models start from the same single-phase checkpoint and use the same PhaseMix-135K splits. We retrain the ablated model for 120,000 phase-stage updates followed by 40,000 STOP-stage updates and evaluate the final STOP checkpoint on all 100,000 test samples. The phase-stage batch sizes are 126 for the full model and 128 for the ablation; both use an effective STOP-stage batch size of 224. The comparison uses one training run per configuration.

Table~\ref{tab:physical-guidance-automatic} reports identification and contribution reconstruction from automatically predicted sets, using the metrics in Appendix~\ref{app:evaluation-metrics}. In this comparison, the full model has 9.0\% lower contribution MAE with Greedy and 11.3\% lower MAE with Beam-10. Exact-set@1 increases by 0.36 and 0.62 percentage points, respectively.

\begin{table}[!htbp]
\centering
\caption{Physics-guidance ablation on PhaseMix-135K. Contribution scores use the first-ranked predicted set.}
\label{tab:physical-guidance-automatic}
\small
\setlength{\tabcolsep}{4pt}
\begin{tabular}{@{}llrrrrr@{}}
\toprule
Model & Search & \multicolumn{2}{c}{Exact-set (\%)} & Phase recall (\%) & \multicolumn{2}{c}{Contribution} \\
\cmidrule(lr){3-4}\cmidrule(l){6-7}
 & & @1 & @10 & & Cosine & MAE ($10^{-4}$) \\
\midrule
Without physics guidance & Greedy & 76.79 & - & 91.37 & 0.9725 & 3.7602 \\
Full model & Greedy & 77.15 & - & 91.50 & 0.9735 & 3.4209 \\
Without physics guidance & Beam-10 & 81.36 & 94.49 & 93.31 & 0.9738 & 3.4133 \\
Full model & Beam-10 & 81.98 & 94.73 & 93.59 & 0.9765 & 3.0280 \\
\bottomrule
\end{tabular}
\end{table}

\paragraph{Stepwise reconstruction with correct phases supplied.}
We apply the correct-phase evaluation in Appendix~\ref{app:stepwise} to both final checkpoints, covering the same 300,000 prefixes. References enter as an ID-sorted set; outputs are mapped back to their identities for scoring. Contribution metrics average over selected phases within each sample, then over samples in each phase-count and step group. Table~\ref{tab:physical-guidance-stepwise} includes the single-phase results omitted from Figure~\ref{fig:physical-guidance-steps}.

Table~\ref{tab:physical-guidance-controlled} aggregates the 200,000 incomplete and 100,000 complete prefixes separately. Physics guidance reduces contribution MAE by 16.4\% for incomplete prefixes and 10.7\% for complete prefixes, with corresponding residual MAE reductions of 21.6\% and 5.2\%. Contribution MAE is lower at every phase-count/step combination; contribution cosine differs only slightly and is not uniformly higher. Residual MAE is lower at all incomplete steps. At the final four-phase step, it instead increases from $4.2540\times10^{-4}$ to $4.4949\times10^{-4}$, while residual cosine is nearly equal. Complete-set residuals retain background and noise rather than serving another true-phase prediction.

\begin{table}[!htbp]
\centering\small
\caption{Correct-phase reconstruction with and without physics guidance, averaged over all test prefixes of each type. MAE is in units of $10^{-4}$.}
\label{tab:physical-guidance-controlled}
\setlength{\tabcolsep}{3pt}
\begin{tabular*}{\linewidth}{@{\extracolsep{\fill}}llrrrr@{}}
\toprule
 & & \multicolumn{2}{c}{Contribution} & \multicolumn{2}{c}{Residual} \\
\cmidrule(lr){3-4}\cmidrule(l){5-6}
History & Physics guidance & Cosine & MAE & Cosine & MAE \\
\midrule
Incomplete & Without & 0.9988 & 5.3124 & 0.9895 & 6.5233 \\
Incomplete & With & 0.9989 & 4.4417 & 0.9928 & 5.1121 \\
Complete & Without & 0.9964 & 3.1938 & 0.9734 & 4.0841 \\
Complete & With & 0.9966 & 2.8534 & 0.9770 & 3.8709 \\
\bottomrule
\end{tabular*}
\end{table}

\begin{table}[!htbp]
\centering\small
\caption{Contribution and residual reconstruction. Metric pairs: without physics guidance (left), full model (right). $K$ denotes phase count; MAE uses units of $10^{-4}$.}
\label{tab:physical-guidance-stepwise}
\setlength{\tabcolsep}{3pt}
\begin{tabular*}{\linewidth}{@{\extracolsep{\fill}}ccrrrr@{}}
\toprule
 & & \multicolumn{2}{c}{Contribution} & \multicolumn{2}{c}{Residual} \\
\cmidrule(lr){3-4}\cmidrule(l){5-6}
$K$ & Step & Cosine & MAE & Cosine & MAE \\
\midrule
1 & 1 & 1.0000 / 1.0000 & 2.7705 / 2.4017 & 0.9836 / 0.9870 & 2.4554 / 2.0867 \\
\midrule
2 & 1 & 0.9997 / 0.9998 & 4.3979 / 3.5523 & 0.9950 / 0.9962 & 4.1862 / 3.3406 \\
2 & 2 & 0.9977 / 0.9977 & 3.2255 / 2.8268 & 0.9750 / 0.9806 & 3.6667 / 3.1297 \\
\midrule
3 & 1 & 0.9995 / 0.9997 & 5.4680 / 4.3406 & 0.9975 / 0.9982 & 5.3233 / 4.1959 \\
3 & 2 & 0.9988 / 0.9987 & 4.5791 / 3.9769 & 0.9861 / 0.9902 & 6.0940 / 4.7667 \\
3 & 3 & 0.9957 / 0.9958 & 3.6703 / 3.2748 & 0.9665 / 0.9735 & 4.6790 / 4.1276 \\
\midrule
4 & 1 & 0.9994 / 0.9995 & 6.3130 / 4.9769 & 0.9983 / 0.9987 & 6.2102 / 4.8740 \\
4 & 2 & 0.9987 / 0.9987 & 5.5393 / 4.6957 & 0.9924 / 0.9949 & 7.8877 / 6.0839 \\
4 & 3 & 0.9975 / 0.9976 & 4.9753 / 4.5215 & 0.9716 / 0.9808 & 7.8627 / 6.2103 \\
4 & 4 & 0.9955 / 0.9958 & 2.9264 / 2.6635 & 0.9752 / 0.9753 & 4.2540 / 4.4949 \\
\bottomrule
\end{tabular*}
\end{table}

\begin{figure}[!htbp]
\centering
\includegraphics{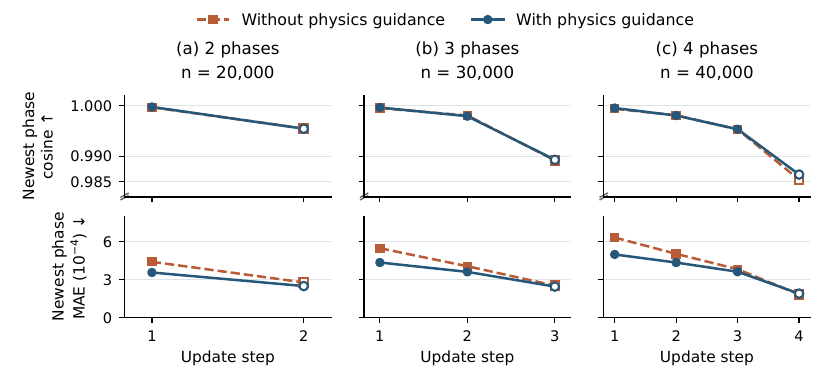}
\caption{Newly added phase contributions with correct phases supplied. (a--c) Two-, three-, and four-phase test samples. Points are sample means; open markers denote the last update. Cosine axes are zoomed.}
\label{fig:physical-guidance-newest}
\end{figure}

\paragraph{Newly added contributions.}
Figure~\ref{fig:physical-guidance-newest} separately scores only the contribution of the phase added at the current step. Its MAE is lower with physics guidance at all incomplete steps; the final four-phase step is slightly worse. Cosine differences are small and mixed. This newest-phase statistic is distinct from the all-selected-phase average used in the main figure.

\FloatBarrier

\subsection{Identification across weak-phase intensity and spectral overlap}
\label{app:weak-overlap}

We stratify the main-table predictions for all 90,000 multiphase PhaseMix-135K test mixtures, comparing PhaseMatcher and Scalar under Greedy with the three external baselines. True contribution spectra define the same groups for all methods; all predictions are included.

\paragraph{Intensity and overlap definitions.}
Let $c_i$ be the weighted, perturbed contribution of true phase $i$ before background and noise are added. We define its integrated intensity share and the weakest phase as
\begin{equation}
s_i=\frac{\sum_{\ell=1}^{L}c_i(\ell)}{\sum_{j=1}^{K}\sum_{\ell=1}^{L}c_j(\ell)},
\qquad i_{\mathrm{w}}=\arg\min_i s_i.
\label{eq:weak-intensity}
\end{equation}
The share measures diffraction intensity, not mass fraction. Spectral overlap measures the cosine similarity between the weakest contribution and the sum of the remaining contributions:
\begin{equation}
o=\frac{c_{i_{\mathrm{w}}}^{\mathsf T}\bigl(\sum_{j\ne i_{\mathrm{w}}}c_j\bigr)}
{\|c_{i_{\mathrm{w}}}\|_2\,\bigl\|\sum_{j\ne i_{\mathrm{w}}}c_j\bigr\|_2}.
\label{eq:weak-overlap}
\end{equation}
Larger $o$ indicates greater similarity to the remaining signal, rather than a percentage of coincident peaks. Weakest-phase recall is the fraction of mixtures for which the weakest true phase ID appears in the predicted set.

\begin{figure}[!htbp]
\centering
\includegraphics{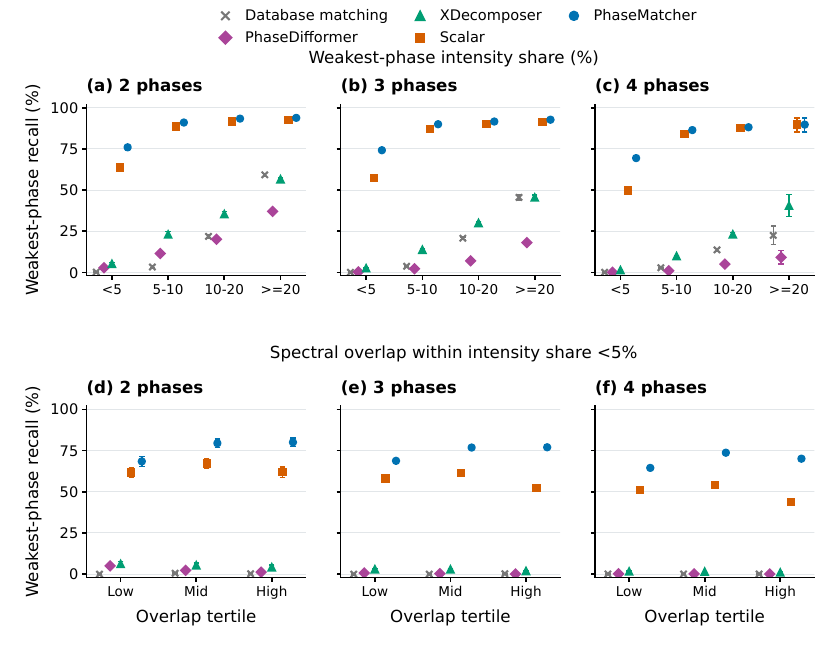}
\caption{\textbf{Weakest-phase identification on PhaseMix-135K.} (a--c) Recall by weakest-phase intensity share. (d--f) Recall by overlap tertile among mixtures with weakest-phase share below 5\%. Columns correspond to two-, three-, and four-phase mixtures. Points show recall with 95\% mixture-bootstrap confidence intervals; some intervals are smaller than the markers. Group sizes appear in Table~\ref{tab:weak-overlap}.}
\label{fig:weak-overlap}
\end{figure}

\paragraph{Grouping and uncertainty.}
Within each phase count, we use intensity-share bins $[0,0.05)$, $[0.05,0.10)$, $[0.10,0.20)$, and $[0.20,1]$. We then divide the weak-phase subset $s_{i_{\mathrm{w}}}<0.05$ into overlap tertiles, independently for each phase count. The two cut points are $(0.02273,0.06274)$, $(0.03164,0.07547)$, and $(0.03903,0.08677)$ for two, three, and four phases, respectively; ``high overlap'' denotes the upper tertile within each subset. Figure~\ref{fig:weak-overlap} reports weakest-phase recall, and Table~\ref{tab:weak-overlap} reports sample counts and Exact-set@1 for the same groups. Intervals are 95\% percentile intervals from 5,000 paired mixture-level bootstrap resamples within each group (seed 20260916).

\paragraph{Results.}
The largest recall gains over Scalar occur below 5\% intensity share: 12.35, 16.89, and 19.75 percentage points for two, three, and four phases. The external baselines also have markedly lower weakest-phase recall, particularly in these low-intensity groups. In four-phase mixtures, the gain over Scalar falls to 2.40 points at 5--10\% and 0.40 points at 10--20\%. Within the below-5\% subset, the advantage increases across overlap tertiles at every phase count. For four phases, weakest-phase recall gains are 13.38, 19.69, and 26.18 points from low to high overlap. In the high-overlap group, Exact-set@1 improves from 37.40\% to 64.17\%. Absolute recall need not decrease monotonically with overlap; the consistent trend is the larger advantage over Scalar. Defining the weakest phase by minimum mixture weight instead gives recall gains of 11.82, 16.53, and 19.51 points below 5\% weight, supporting the same low-intensity trend.

\begin{table}[!htbp]
\centering\small
\caption{Exact-set@1 (\%) in the groups of Figure~\ref{fig:weak-overlap}. Database denotes iterative database matching and PM denotes PhaseMatcher. The two PhaseMatcher variants and database matching use Greedy.}
\label{tab:weak-overlap}
\setlength{\tabcolsep}{2pt}
\begin{tabular*}{\linewidth}{@{\extracolsep{\fill}}llrrrrrr@{}}
\toprule
$K$ & Group & $n$ & Database & PhaseDifformer & XDecomposer & Scalar & PM \\
\midrule
\multicolumn{8}{l}{(a) Weakest-phase intensity share} \\
2 & $<5\%$ & 2,688 & 0.26 & 2.42 & 4.32 & 62.39 & 75.37 \\
 & $5$--$10\%$ & 2,353 & 2.85 & 8.50 & 18.49 & 85.81 & 88.65 \\
 & $10$--$20\%$ & 4,098 & 17.86 & 11.01 & 26.96 & 87.92 & 90.04 \\
 & $\geq20\%$ & 10,861 & 41.26 & 7.71 & 38.54 & 87.63 & 89.55 \\
\addlinespace[3pt]
3 & $<5\%$ & 10,614 & 0.10 & 0.05 & 0.63 & 51.74 & 69.52 \\
 & $5$--$10\%$ & 7,150 & 2.73 & 0.11 & 3.86 & 77.69 & 82.21 \\
 & $10$--$20\%$ & 8,573 & 13.82 & 0.51 & 8.92 & 79.66 & 82.33 \\
 & $\geq20\%$ & 3,663 & 29.13 & 0.71 & 12.56 & 79.42 & 81.79 \\
\addlinespace[3pt]
4 & $<5\%$ & 23,303 & 0.09 & 0.00 & 0.00 & 40.75 & 61.98 \\
 & $5$--$10\%$ & 10,480 & 1.76 & 0.00 & 0.00 & 68.37 & 74.55 \\
 & $10$--$20\%$ & 6,022 & 7.54 & 0.03 & 0.00 & 69.66 & 73.33 \\
 & $\geq20\%$ & 195 & 11.28 & 0.00 & 0.00 & 73.85 & 73.85 \\
\midrule
\multicolumn{8}{l}{(b) Overlap tertiles within intensity share $<5\%$} \\
2 & Low & 896 & 0.00 & 4.35 & 4.46 & 59.15 & 66.96 \\
 & Mid & 896 & 0.56 & 1.79 & 4.91 & 66.85 & 79.24 \\
 & High & 896 & 0.22 & 1.12 & 3.57 & 61.16 & 79.91 \\
\addlinespace[3pt]
3 & Low & 3,538 & 0.00 & 0.06 & 0.71 & 51.16 & 63.43 \\
 & Mid & 3,538 & 0.06 & 0.06 & 0.73 & 56.25 & 72.36 \\
 & High & 3,538 & 0.25 & 0.03 & 0.45 & 47.82 & 72.78 \\
\addlinespace[3pt]
4 & Low & 7,768 & 0.10 & 0.01 & 0.00 & 39.97 & 55.45 \\
 & Mid & 7,767 & 0.06 & 0.00 & 0.00 & 44.88 & 66.33 \\
 & High & 7,768 & 0.10 & 0.00 & 0.00 & 37.40 & 64.17 \\
\bottomrule
\end{tabular*}
\end{table}

\FloatBarrier

\subsection{Greedy inference efficiency}
\label{app:runtime}

We measure Greedy inference on a single RTX 4090 with BF16 mixed precision over all 135,258 candidate records, using 256 test spectra for each phase count from one to four. Table~\ref{tab:greedy-runtime} reports mean single-input inference time, excluding model and data loading.

\begin{table}[H]
\centering\small
\caption{Single-GPU Greedy inference time on PhaseMix-135K.}
\label{tab:greedy-runtime}
\begin{tabular*}{\linewidth}{@{\extracolsep{\fill}}lrrrr@{}}
\toprule
Phase count & 1 & 2 & 3 & 4 \\
\midrule
Time (s/mixture) & 0.0250 & 0.0401 & 0.0551 & 0.0685 \\
\bottomrule
\end{tabular*}
\end{table}

\FloatBarrier
\clearpage
\subsection{RRUFF contribution examples}
\label{app:rruff-cases}
\newcommand{\rruffcasegraphic}[1]{\makebox[\linewidth][l]{\hspace*{-0.22in}\includegraphics[width=5.84in]{#1}}}

\begin{figure}[H]
\centering
\rruffcasegraphic{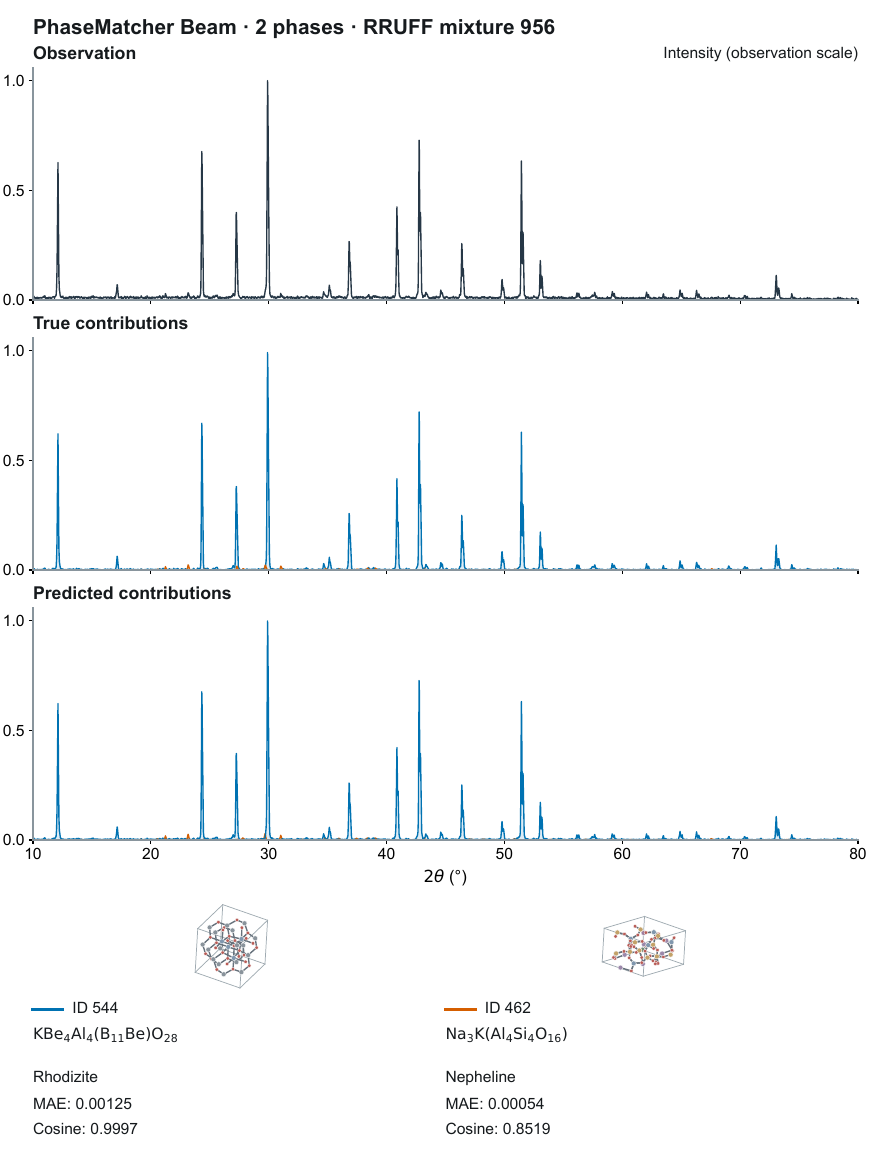}
\caption{Two-phase RRUFF-based mixture 956. Weakest-phase share: 3.39\%; overlap: 0.108.}
\label{fig:rruff-beam-two}
\end{figure}

\begin{figure}[H]
\centering
\rruffcasegraphic{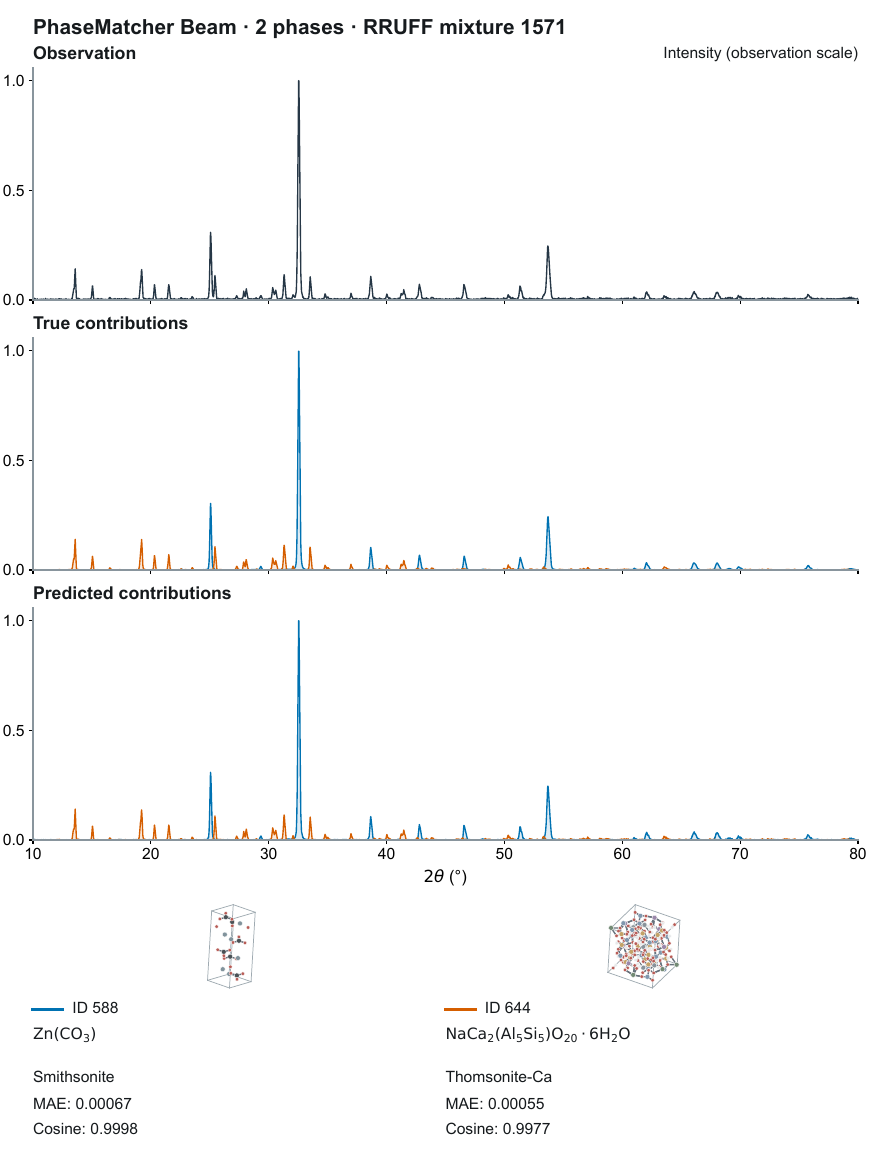}
\caption{Two-phase RRUFF-based mixture 1571. Weakest-phase share: 35.31\%; overlap: 0.013.}
\label{fig:rruff-beam-1571}
\end{figure}

\begin{figure}[H]
\centering
\rruffcasegraphic{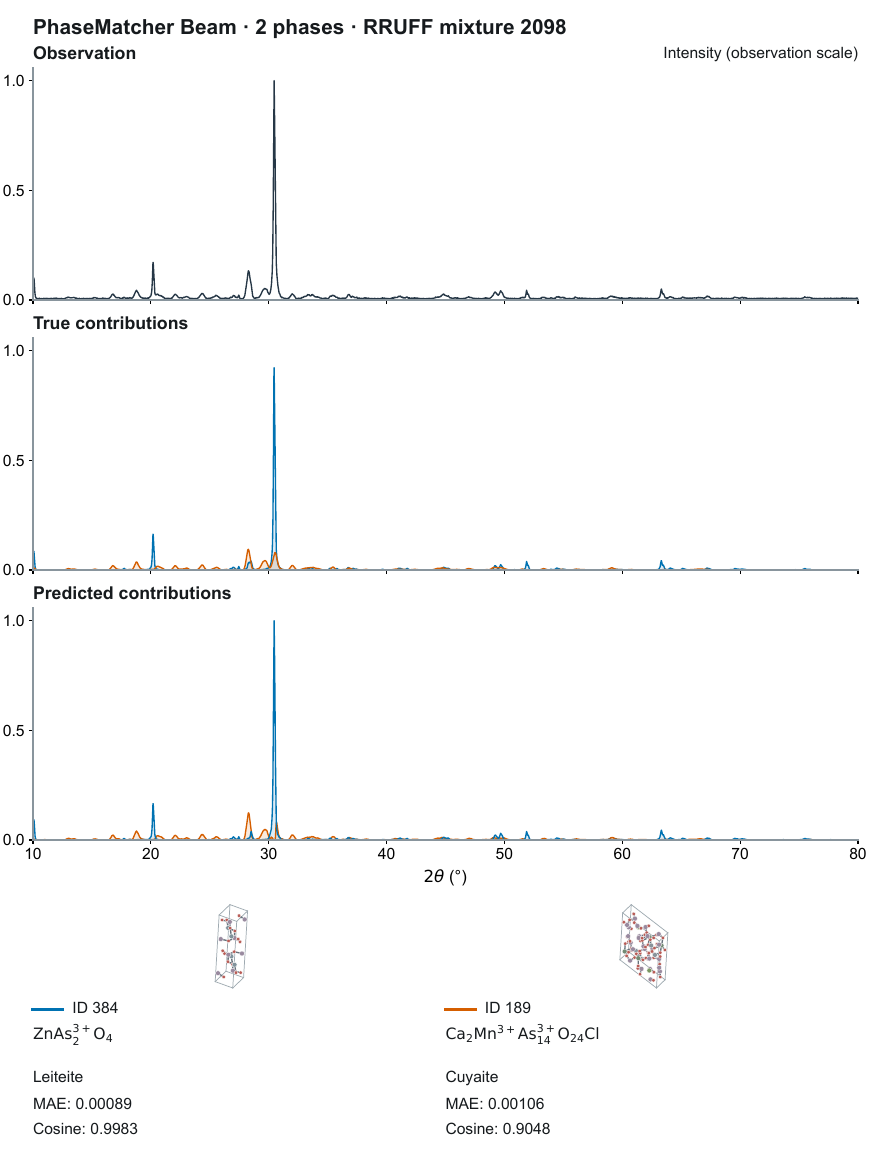}
\caption{Two-phase RRUFF-based mixture 2098. Weakest-phase share: 43.63\%; overlap: 0.481.}
\label{fig:rruff-beam-2098}
\end{figure}

\begin{figure}[H]
\centering
\rruffcasegraphic{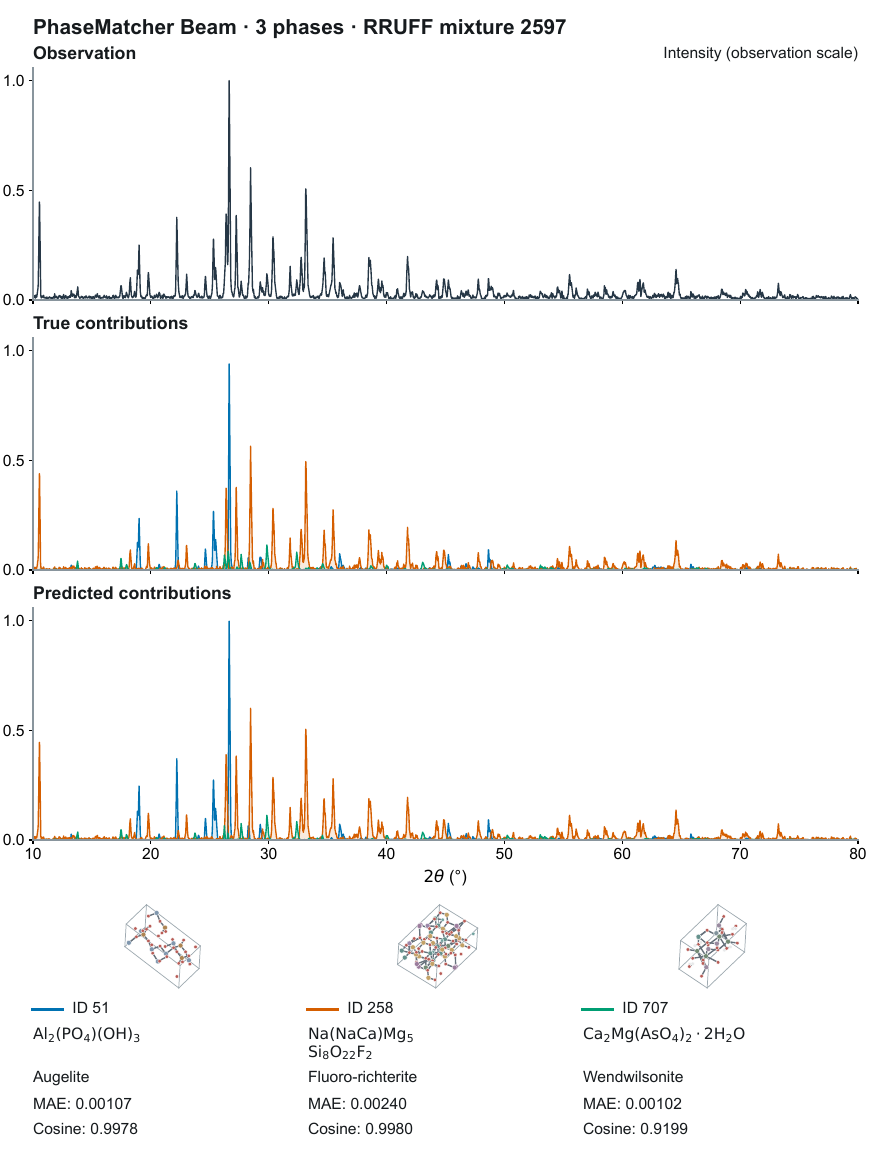}
\caption{Three-phase RRUFF-based mixture 2597. Weakest-phase share: 11.02\%; overlap: 0.286.}
\label{fig:rruff-beam-three}
\end{figure}

\begin{figure}[H]
\centering
\rruffcasegraphic{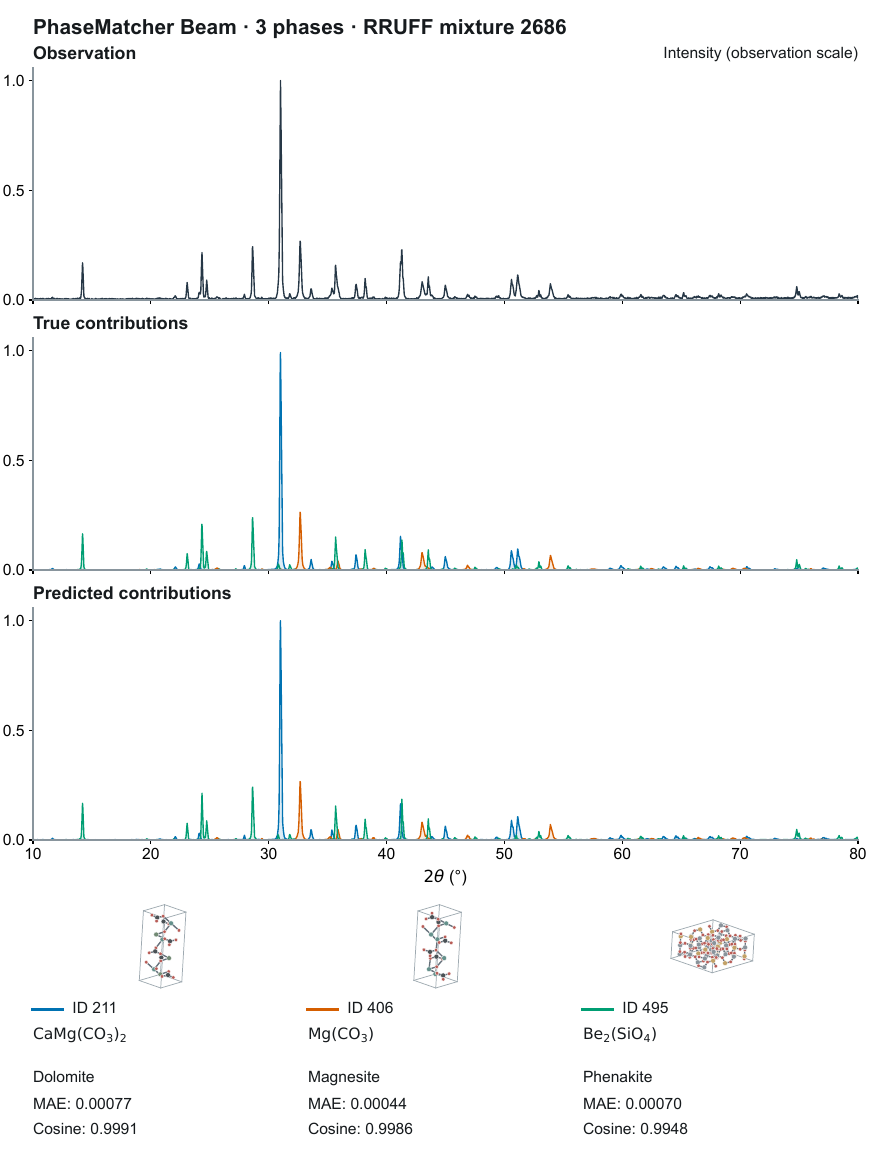}
\caption{Three-phase RRUFF-based mixture 2686. Weakest-phase share: 20.10\%; overlap: 0.033.}
\label{fig:rruff-beam-2686}
\end{figure}

\begin{figure}[H]
\centering
\rruffcasegraphic{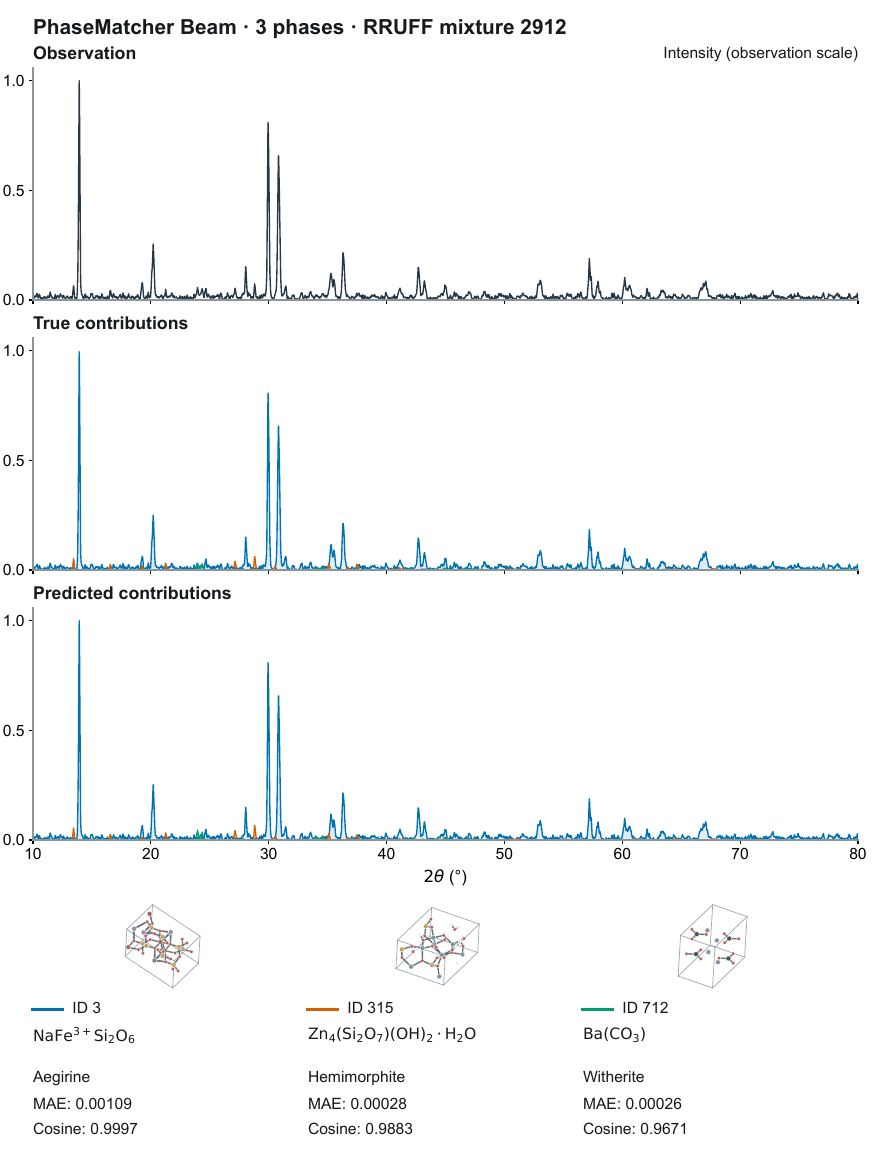}
\caption{Three-phase RRUFF-based mixture 2912. Weakest-phase share: 2.97\%; overlap: 0.081.}
\label{fig:rruff-beam-2912}
\end{figure}

\begin{figure}[H]
\centering
\rruffcasegraphic{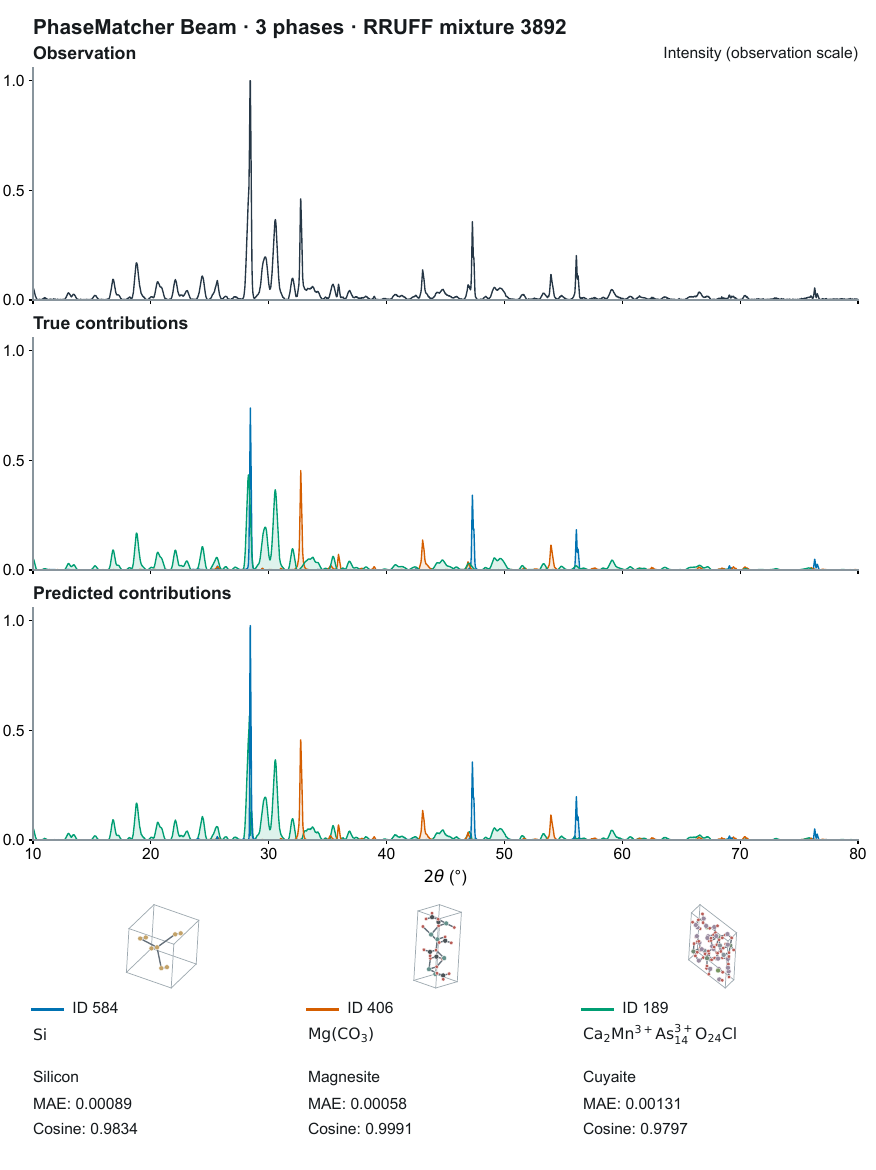}
\caption{Three-phase RRUFF-based mixture 3892. Weakest-phase share: 15.40\%; overlap: 0.264.}
\label{fig:rruff-beam-3892}
\end{figure}

\begin{figure}[H]
\centering
\rruffcasegraphic{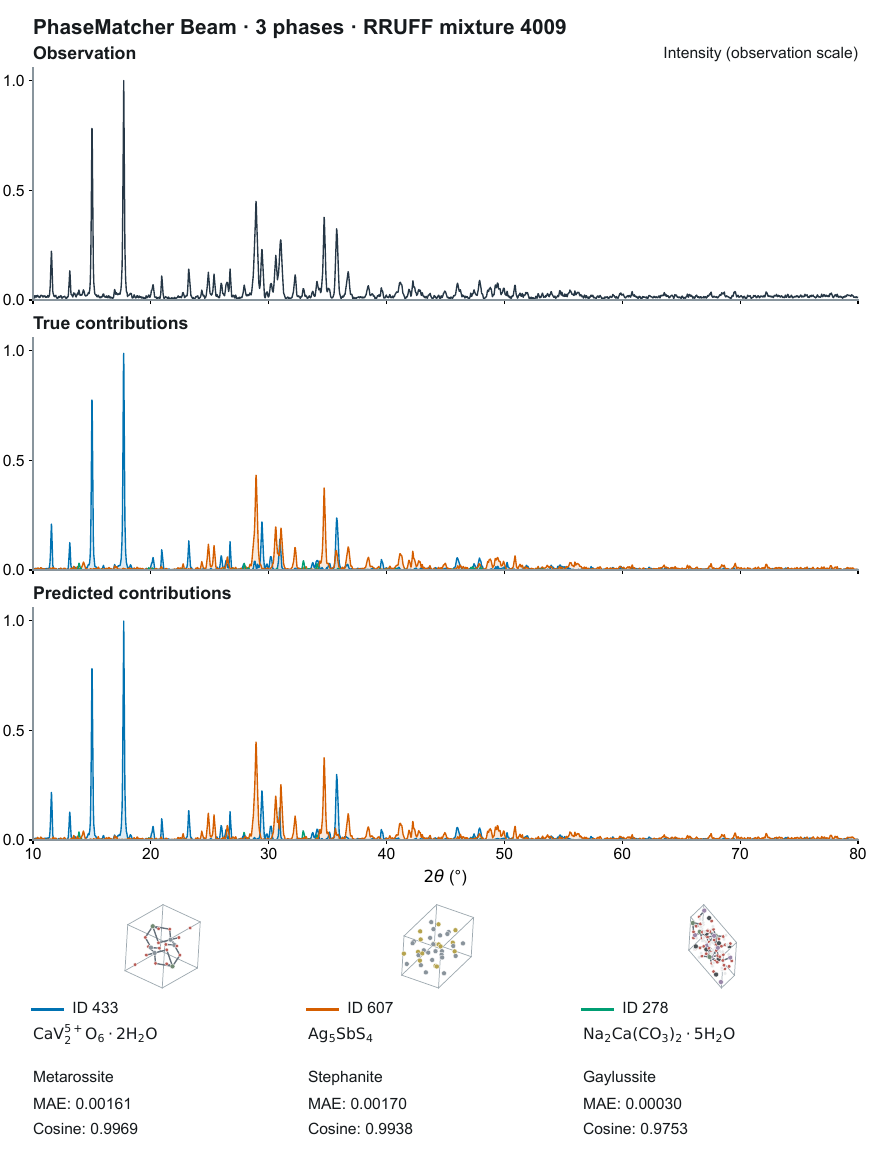}
\caption{Three-phase RRUFF-based mixture 4009. Weakest-phase share: 2.97\%; overlap: 0.122.}
\label{fig:rruff-beam-4009}
\end{figure}

\begin{figure}[H]
\centering
\rruffcasegraphic{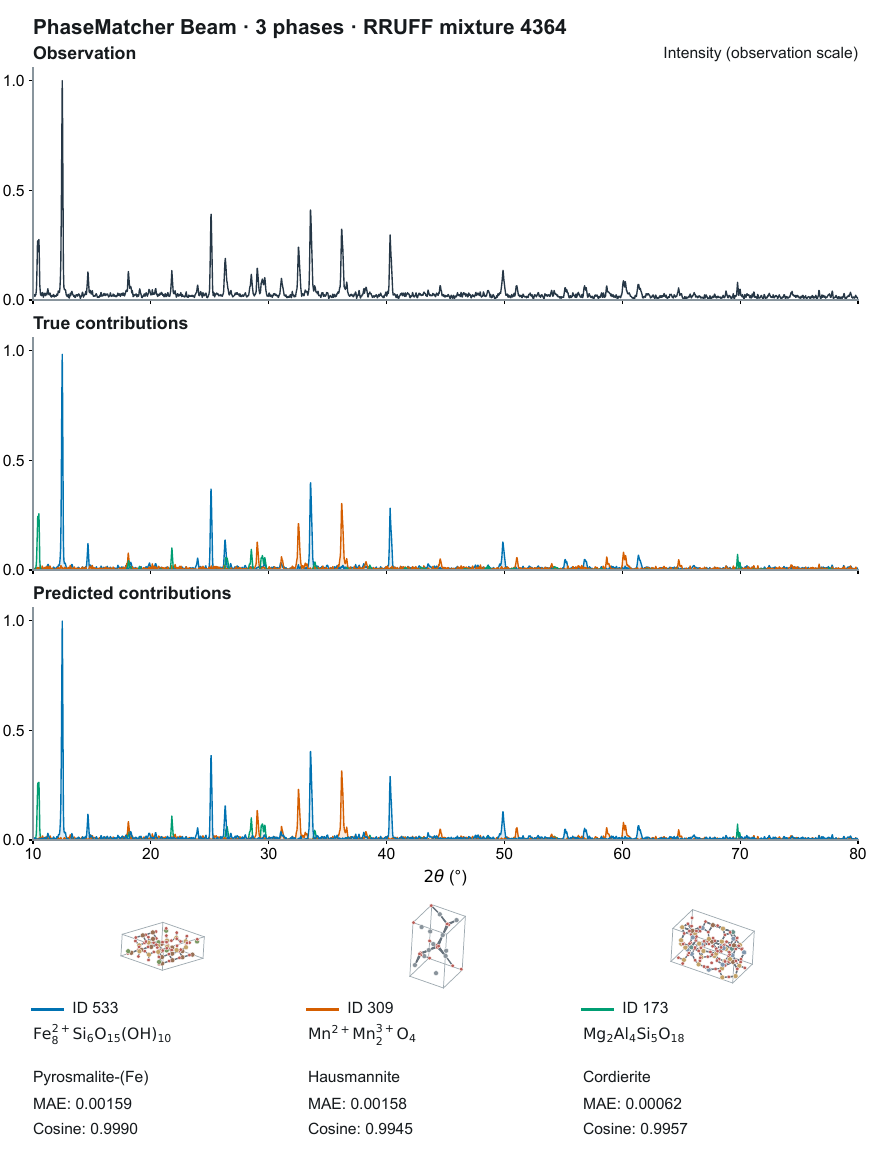}
\caption{Three-phase RRUFF-based mixture 4364. Weakest-phase share: 11.74\%; overlap: 0.086.}
\label{fig:rruff-beam-4364}
\end{figure}

\begin{figure}[H]
\centering
\rruffcasegraphic{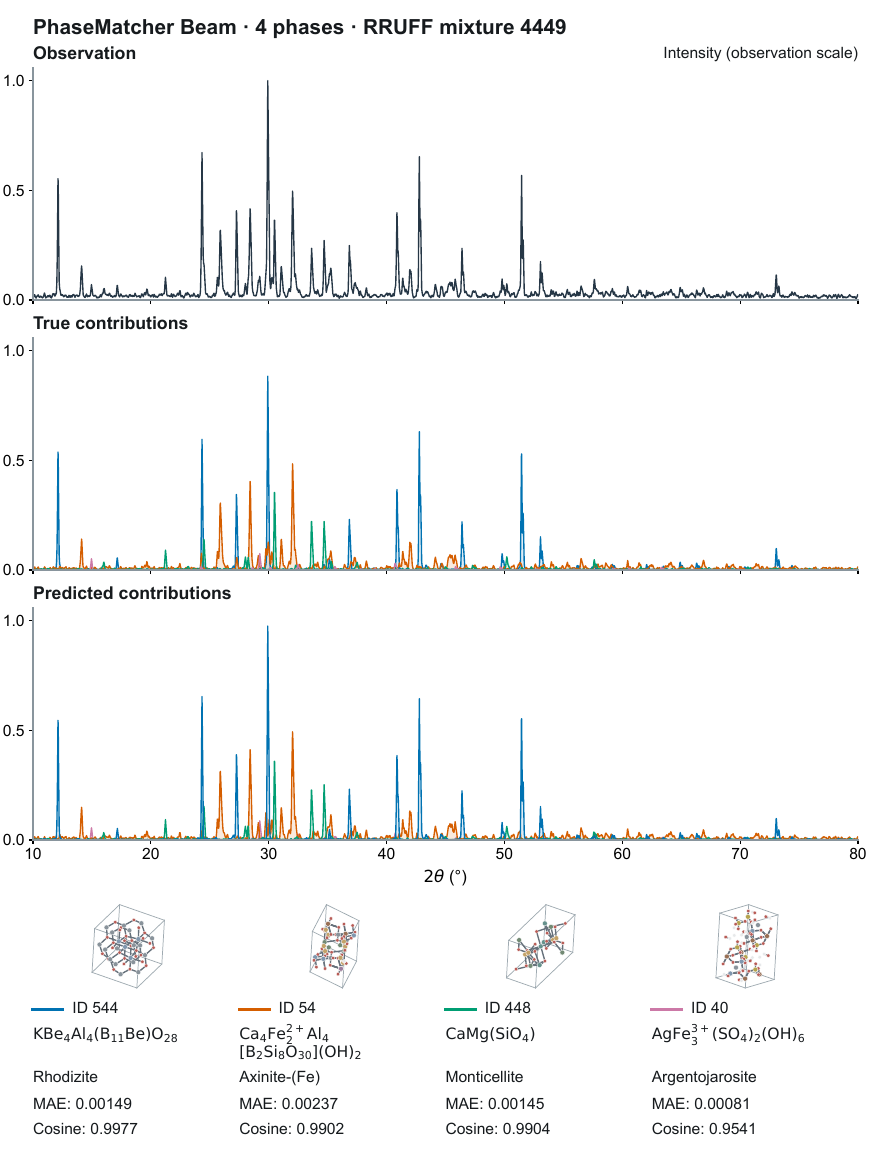}
\caption{Four-phase RRUFF-based mixture 4449. Weakest-phase share: 3.84\%; overlap: 0.168.}
\label{fig:rruff-beam-four}
\end{figure}

\begin{figure}[H]
\centering
\rruffcasegraphic{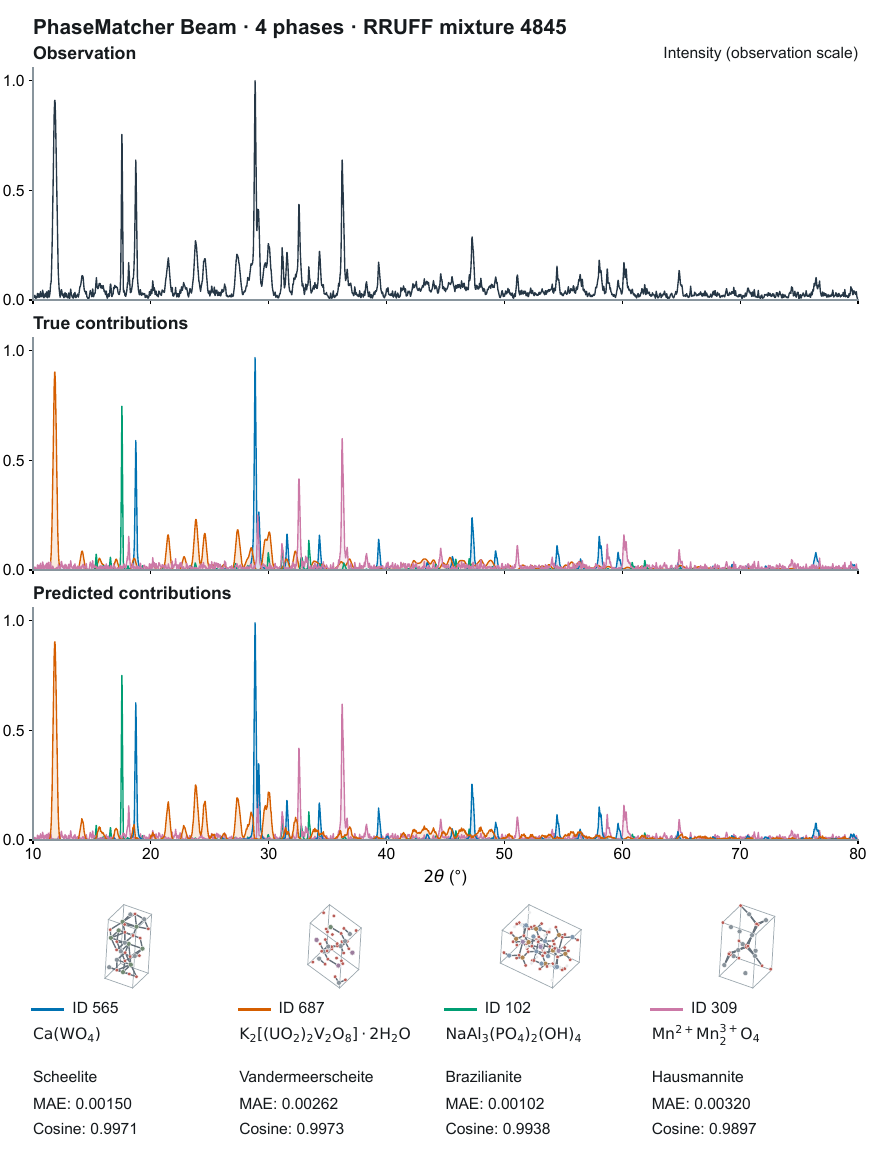}
\caption{Four-phase RRUFF-based mixture 4845. Weakest-phase share: 7.95\%; overlap: 0.077.}
\label{fig:rruff-beam-4845}
\end{figure}

\begin{figure}[H]
\centering
\rruffcasegraphic{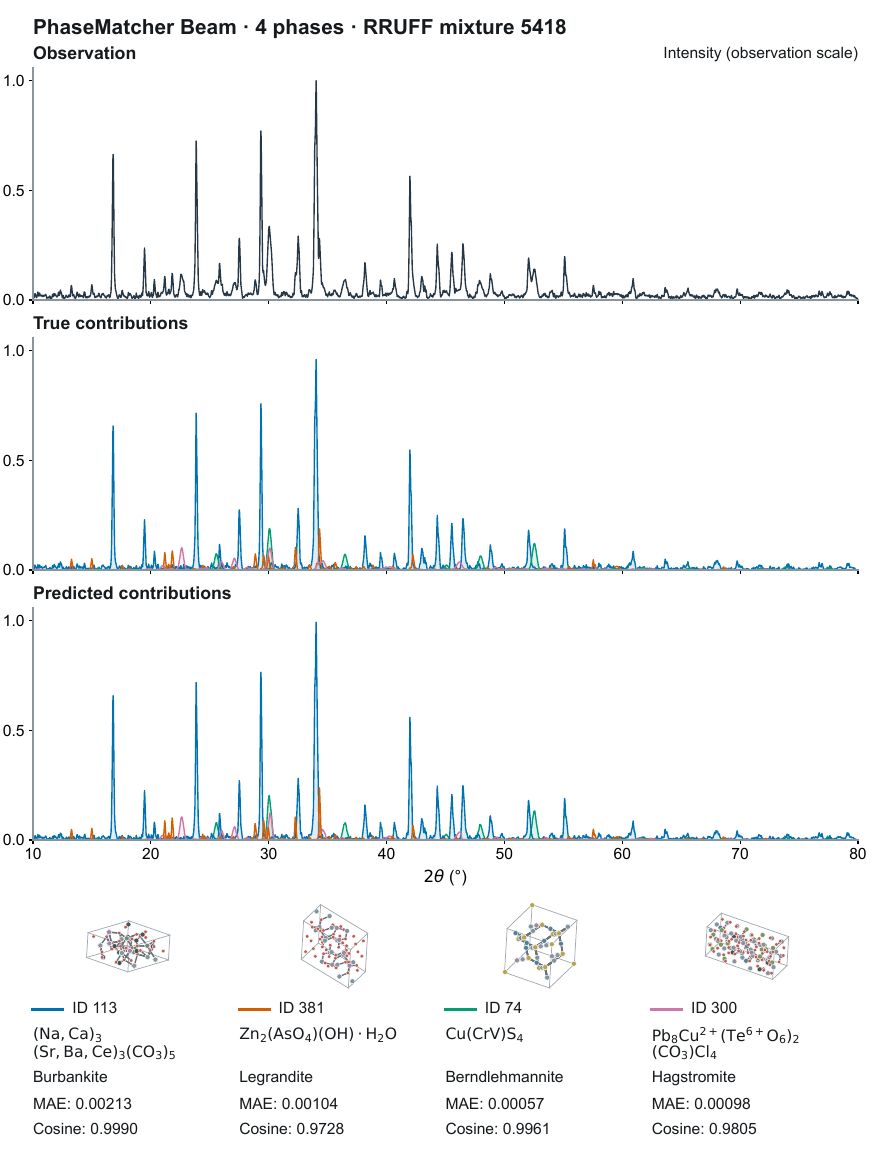}
\caption{Four-phase RRUFF-based mixture 5418. Weakest-phase share: 9.34\%; overlap: 0.112.}
\label{fig:rruff-beam-5418}
\end{figure}

\begin{figure}[H]
\centering
\rruffcasegraphic{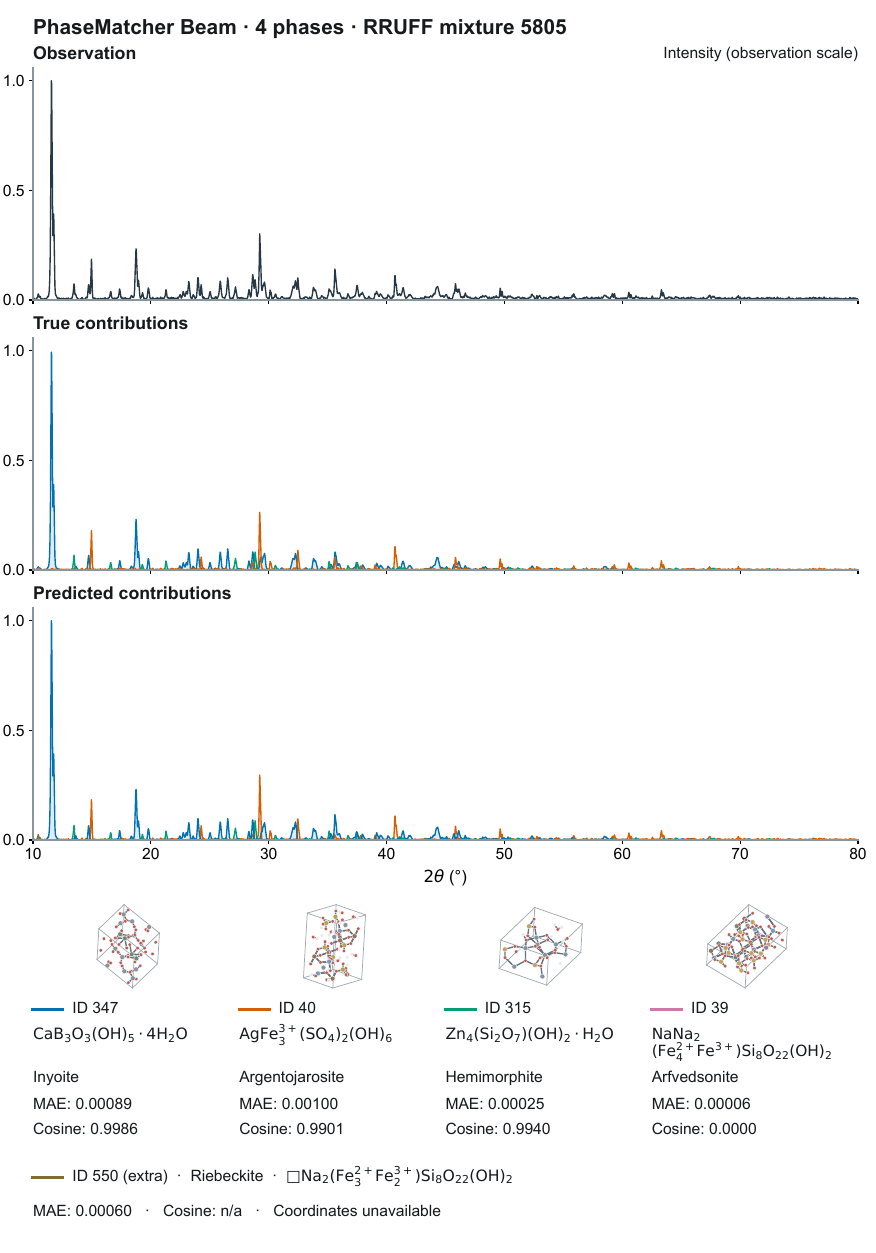}
\caption{Four-phase RRUFF-based mixture 5805. Weakest-phase share: 0.34\%; overlap: 0.080. Beam misses ID 39 and adds ID 550.}
\label{fig:rruff-beam-5805}
\end{figure}

\begin{figure}[H]
\centering
\rruffcasegraphic{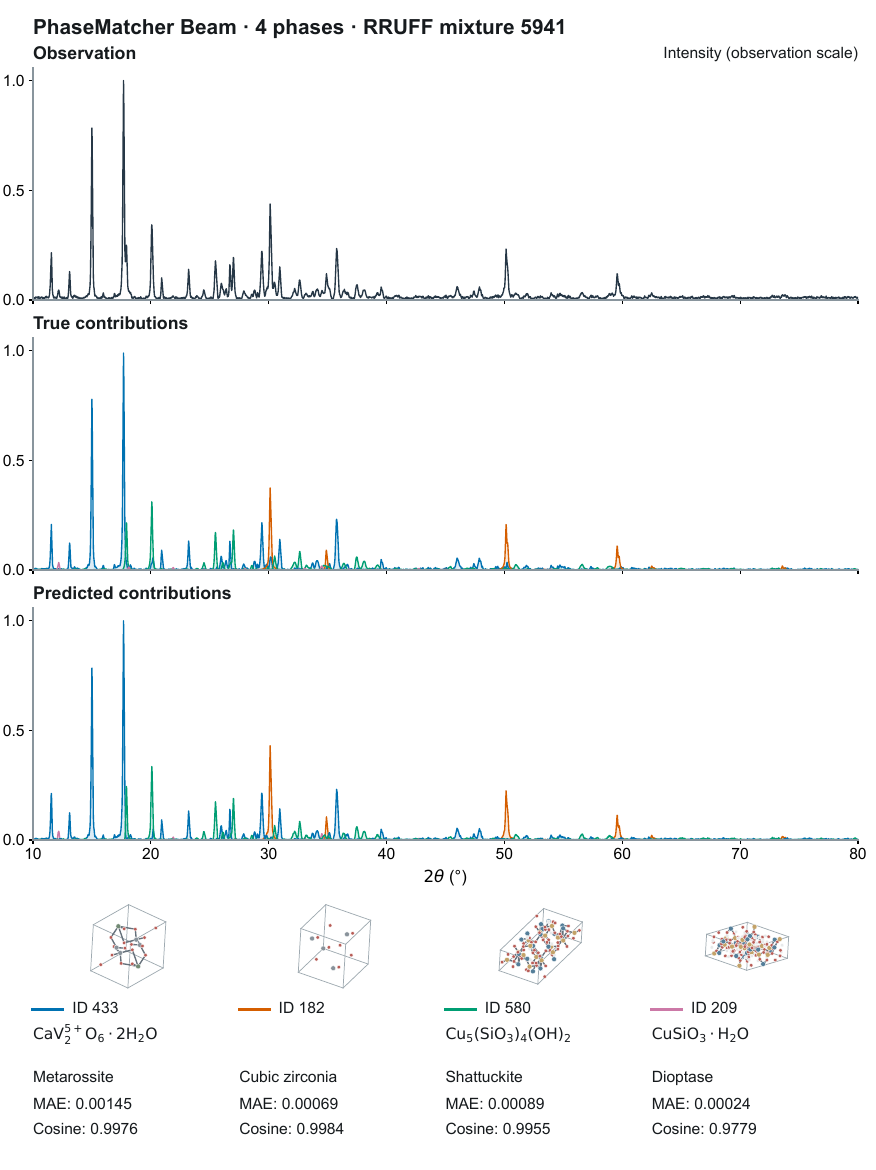}
\caption{Four-phase RRUFF-based mixture 5941. Weakest-phase share: 2.16\%; overlap: 0.071.}
\label{fig:rruff-beam-5941}
\end{figure}

\begin{figure}[H]
\centering
\rruffcasegraphic{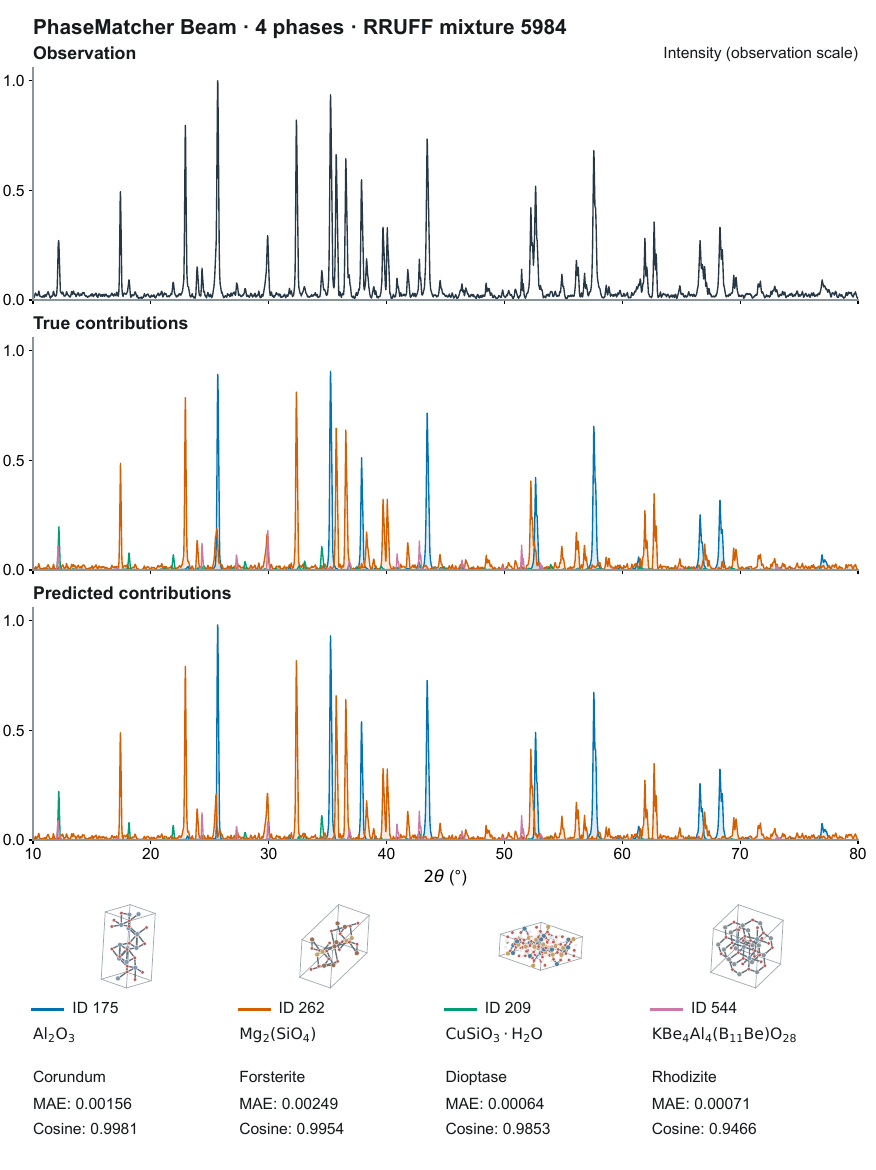}
\caption{Four-phase RRUFF-based mixture 5984. Weakest-phase share: 4.57\%; overlap: 0.102.}
\label{fig:rruff-beam-5984}
\end{figure}

Figures~\ref{fig:rruff-beam-two}--\ref{fig:rruff-beam-5984} show PhaseMatcher Beam on 15 controlled mixtures constructed from single-phase experimental RRUFF patterns: three two-phase, six three-phase, and six four-phase examples. We retain the original three examples and select 12 more to cover different weak-phase intensity shares and spectral overlaps among mixtures with available atomic coordinates, without using prediction outcomes. Each figure shows the observation, true contributions, and final Beam contributions on the same intensity scale.

Colors, chemical formulas, and library IDs identify phases; crystal drawings use available DIF-file structures. Per-phase MAE and cosine compare predicted and true contributions over the full angular range. We assign zero predictions and cosine zero to missing phases, and compare extra predictions with a zero target (cosine undefined). Captions report the weakest phase's integrated-intensity share and its cosine overlap with the sum of the other true contributions.

\clearpage

\clearpage

\clearpage

\clearpage

\clearpage

\clearpage

\clearpage

\clearpage

\clearpage

\clearpage

\clearpage

\clearpage

\clearpage

\clearpage

\FloatBarrier

\end{document}